\RequirePackage[l2tabu, orthodox]{nag}
\RequirePackage{snapshot}

\documentclass[9pt,onecolumn]{extarticle}

\makeatletter
\if@twocolumn
  \usepackage[dvips,letterpaper,top=0.5in, bottom=0.5in, left=0.75in, right=0.5in,includefoot,heightrounded]{geometry}
\else
  \usepackage[dvips,letterpaper,margin=1in,includefoot,heightrounded]{geometry}
\fi

\usepackage{srcltx}

\usepackage[russian,portuges,english]{babel}

\iflanguage{portuges}
    {\newcommand{\keywordname}{Palavras-chaves}}
    {\newcommand{\keywordname}{Keywords}}

\usepackage{amsmath}
\usepackage{amssymb}
\usepackage{amsfonts}
\usepackage{mathtools}

\usepackage{abstract}

\usepackage{graphicx}
\usepackage{subfig}
\usepackage[usenames,dvipsnames,svgnames,x11names]{xcolor}

\usepackage{booktabs}
\usepackage{arydshln}

\usepackage{setspace}
\usepackage{flushend}

\usepackage{multicol}

\usepackage{cite}

\usepackage{hyperref}
\usepackage[normalem]{ulem}

\usepackage{enumerate}

\usepackage{algorithm}
\usepackage{algpseudocode}

\usepackage{psfrag}
\usepackage{multirow}
\usepackage[squaren]{SIunits}

\usepackage{listings}

\usepackage[short,12hr]{datetime}
       \usepackage{fouriernc}
\makeatletter

\newenvironment{rsmallmatrix}{\null\,\vcenter\bgroup
  \Let@\restore@math@cr\default@tag
  \baselineskip6\ex@ \lineskip1.5\ex@ \lineskiplimit\lineskip
  \ialign\bgroup\hfil$\m@th\scriptstyle##$&&\thickspace\hfil
  $\m@th\scriptstyle##$\crcr
}{%
  \crcr\egroup\egroup\,%
}

\newcommand{\printtitle}{%
\makeatletter
\if@twocolumn

\twocolumn[%
  \maketitle
  \begin{onecolabstract}
    \myabstract
  \end{onecolabstract}
  \begin{center}
    \small
    \textbf{\keywordname}
    \\\medskip
    \mykeywords
  \end{center}
  \bigskip
]
\saythanks
\else
  \maketitle
  \begin{onecolabstract}
    \myabstract
  \end{onecolabstract}
  \begin{center}
    \small
    \textbf{\keywordname}
    \\\medskip
    \mykeywords
  \end{center}
  \bigskip
  \onehalfspacing
\fi
\makeatother
}

\author{%
A. P. Rad\"unz%
\thanks{%
Graduate Program in Statistics,
Universidade Federal de Pernambuco, Recife, Brazil.}
\and
D. F. G. Coelho%
\thanks{Independent researcher, Calgary, Canada.}
\and
F. M. Bayer%
\thanks{Departamento de Estat\'istica and LACESM,
Universidade Federal de Santa Maria,
Santa Maria, 97105-900, Brazil}
\and
R.~J.~Cintra%
\thanks{%
Signal Processing Group,
Department of Technology,
Universidade Federal de Pernambuco, Caruaru, Brazil.
E-mail: \url{rjdsc@de.ufpe.br}}
\and
A. Madanayake%
\thanks{
Department of Electrical and Computer Engineering,
Florida International University, FL, 33174, USA.}
}

\title{%
Fast Data-independent KLT Approximations
Based on Integer Functions}

\newcommand{\myabstract}{%
The Karhunen-Loève transform (KLT) stands as a well-established discrete transform, demonstrating optimal characteristics in data decorrelation and dimensionality reduction. Its ability to condense energy compression into a select few main components has rendered it instrumental in various applications within image compression frameworks.
However, computing the KLT depends on the covariance
	matrix of the input data, which makes it difficult to develop fast algorithms for its implementation.
 Approximations for the KLT, utilizing specific rounding functions, have been introduced to reduce its computational complexity. Therefore, our paper introduces a category of low-complexity, data-independent KLT approximations, employing a range of round-off functions.
The design methodology of the approximate transform is defined for any block-length $N$, but emphasis is given to transforms of $N = 8$ due to its wide use in image and video compression.
 The proposed transforms perform well when
	compared to the exact KLT and approximations
	considering classical performance
	measures. For particular scenarios, our proposed transforms demonstrated superior performance when compared to KLT approximations documented in the literature.
	We also developed fast algorithms for the proposed transforms, further reducing the arithmetic cost associated with their implementation. Evaluation of field programmable gate array (FPGA) hardware implementation metrics was conducted. %
 Practical applications in image encoding showed the relevance
	of the proposed transforms. In fact, we showed that one of the proposed
	transforms outperformed the exact KLT given certain compression ratios.
}

\newcommand{\mykeywords}{%
Approximate transforms, fast algorithms, image compression, Karhunen-Lo\`eve Transform, low-complexity transforms.
}

\date{}

\begin{document}

\printtitle

\section{Introduction}
Among the discrete transforms,
the Karhunen-Lo\`eve transform (KLT) is an optimal linear tool for
data decorrelation,
being capable
of concentrating the signal energy in few
transform-domain coefficients~\cite{ochoa2019discrete}.
Because the KLT
minimizes the mean square error of compressed data,
it can greatly reduce data dimensionality
and
can be regarded as the ideal
transform for image compression~\cite{ochoa2019discrete,britanak2010discrete}.
Despite such good properties,
the KLT finds few practical applications~\cite{jain1976fast,reed1994fast,jayakumar2020karhunen,geetha2020hybrid,zhang2020data,yang2020blind},
mainly due to the fact
that the definition of the KLT matrix
relies
on the covariance matrix of the input signal.
Therefore,
in general,
two different input signals
would
effect
two different transformation matrices,
thus, precluding the design of efficient approaches
for computing the transformed signal.
Moreover,
because of the data dependency,
generally,
the KLT
suffers
from the dictionary exchange problem~\cite{lan1993fast,ochoa2019discrete},
i.e.,
the transformation basis is not known \emph{a priori}
by the decoder.
Considering the relevance of KLT in data compression context and the associated implementation costs, our goal is to introduce low-complexity approximations for the KLT that are independent of the input data.
For specific classes of signals,
such as
first-order Markovian processes with known correlation coefficient $\rho$,
the KLT matrix
can be expressed simply in terms of $\rho$~\cite{ray1970further}.
Nevertheless,
the complexity of
the resulting transformation matrix
remains in $\mathcal{O}(N^2)$,
where $N$ is the signal length~\cite{ochoa2019discrete}
and, in general, there are no efficient fast algorithms available for its computation~\cite{jain1976fast,lan1994improved,chen2012new,thomakos2016smoothing,blahut2010fast}.
In this context, several KLT approximations~\cite{lan1993fast,lan1994improved,pirooz1998new,cagnazzo2006low,wongsawat2006integer,bhairannawar2020fpga} and fast algorithms for the KLT~\cite{jain1976fast,yanyun2004updating,sole2009joint,blanes2012divide,hao2003reversible} have been proposed
so they have a lower computational cost. However, these methods still suffer the problem of data-dependency.

For the particular -- but very relevant -- case
where $\rho\to1$,
the KLT
assumes the mathematical definition of the DCT~\cite{ahmed1974discrete}.
In other words, the DCT is the KLT for
highly autocorrelated
first-order Markovian  data %
\cite{britanak2010discrete,rao1990yip,ochoa2019discrete}.
Such model
fairly captures the structure of natural images,
which is typically assumed to admit $\rho = 0.95$~\cite{gonzalez2002digital}.
Being fully independent of the data~\cite{ochoa2019discrete,britanak2010discrete},
efficient methods for computing the DCT can be derived~\cite{chen1977fast,loeffler1989practical},
such as the Loeffler algorithm and the Chen algorithm,
turning
it
into a central tool for image and video coding~\cite{wallace1992jpeg,Puri2004,pourazad2012hevc}.
However,
even at the reduced computational cost offered
by fast algorithms,
the residual complexity
of the DCT might still
be sufficiently large
to preclude its application
in contexts
where
severe restrictions
on
computational processing power
and/or
energy autonomy
are present,
such as in wireless and satellite communication systems and in portable computing
applications~\cite{britanak2010discrete}.
This rea\-li\-ty opened the path
for the design of extremely low-complexity
methods for the DCT estimation
based on approximate integer transforms.
Hence,
several approximations for the DCT have been proposed,
generally
being multiplierless transforms
that require addition and bit-shifting operations only~\cite{cintra2014low,bouguezel2008low,haweel2001new,cintra2011dct,potluri2014improved,bayer2012dct,jridi2015generalized,da2017multiplierless,oliveira2019low,canterle2020multiparametric,singhadia2019novel,puchala2021approximate,zidani2019low,chen2019hardware,huang2019deterministic,coelho2021scaling,brahimi2020novel,radunz2023extensions,mefoued2023improving,Coutinho2017}.
In particular, we cite the following approximations for the DCT based on integer functions: the signed DCT (SDCT)~\cite{haweel2001new}, the rounded DCT (RDCT)~\cite{cintra2011dct}, and the collection of integer DCT approximations detailed in~\cite{cintra2014low}.
Despite the very low computational requirements,
such approximations can still
offer good coding performance and
constitute realistic alternatives to the exact DCT.

In~\cite{radunz2021low,radunz2021data}, the approximation design based on integer functions
employed in the derivation of the SDCT~\cite{haweel2001new} and RDCT~\cite{cintra2011dct} were extended
to obtain data-independent KLT approximations.
The signed KLT approximations (SKLT)~\cite{radunz2021data} were formulated by employing the signum function on the elements of the exact KLT matrix considering various block-lengths. On the other hand, the rounded KLT (RKLT)~\cite{radunz2021low} were derived by applying a rounding function to the elements of the exact KLT.
These new low-complexity KLT approximations were submitted to
experiments on image and video coding and showed good performance at
low implementation costs.
Given that these KLT approximations were derived for specific rounding function cases, we opted to expand our testing to a broader array of functions, aiming to achieve improved results.
Considering the above discussion
and taking into account
the following major aspects:
\begin{itemize}
	\item
	the current literature lacks
	specific methodologies
	for the low-complexity computation of the KLT, mainly when considering low-complexity approximation transforms for mid- and low-correlated signals;

	\item
	the methods for the KLT evaluation~\cite{jain1976fast,reed1994fast,lan1993fast,lan1994improved,pirooz1998new,yanyun2004updating,cagnazzo2006low,sole2009joint,blanes2012divide,hao2003reversible,wongsawat2006integer,fan2019signal,biswas2010improved}
	exhibit data-dependency
	which entail
	severe difficulties
	in designing
	fast algorithms based on
	matrix factorization;

	\item
	the dictionary exchange problem presented in the KLT and its approximations~\cite{ochoa2019discrete,britanak2010discrete};

	\item
	the proven success of
	matrix approximation theory
	for deriving low-com\-plex\-ity DCT methods found in the literature;

	\item and the fact that the KLT is the optimal linear transform in terms of decorrelation of first-order Markovian signals;
\end{itemize}
we
aim at proposing
approximations
for the KLT with the following properties:
(i)~data independence and closed-form expression;
(ii)~symmetrical structure that leads to sparse
matrix factorizations and fast algorithms,
and
(iii)~suitability for first-order Markovian
processes
at a wide range of correlation coefficient.
To obtain the sought KLT approximations, we consider integer-based
approximation methods as in~\cite{haweel2001new,cintra2011dct,cintra2014low,radunz2021data,radunz2021low}.
Thus, the main goal of our paper is {to propose low-complexity approximate transforms for the KLT considering different values of the correlation coefficients~$\rho$, so low-correlated signals could be properly
	treated as well.}
To the best of our knowledge, the methodology employed to derive these novel approximations is unprecedented in the literature, particularly concerning the application of KLT in first-order Markovian processes. Our objective is to propose low-complexity KLT approximations adaptable to various contexts.

This paper is structured as follows. In Section~\ref{S:KLT}, we present the mathematical formulation of the KLT for first-order Markovian data, a brief review of approximation theory for discrete transforms,  and the assessment metrics used for the evaluation of approximate transforms.
Section~\ref{S:ProposedTransform} introduces the optimization problem, search space, objective function, and the methodology used to obtain the proposed transforms.
The proposed transforms are presented in Section~\ref{S:results} and the fast algorithms
and
their computational complexities are displayed in Section~\ref{S:fastalgo}.
Section~\ref{s:imagecomp} presents the experiments on image compression.
In Section~\ref{sec:hardware} a field-programmable gate array (FPGA) design is proposed and compared with competing methods.
Finally, Section~\ref{S:conclusion} concludes the paper.

\section{KLT and Approximate Transforms}

\label{S:KLT}
\subsection{KLT for First-Order Markovian Signal}

The KLT
maps
an
$N$-point input signal
$\mathbf{x} =
\begin{bmatrix}
	x_0 & x_1 & \ldots & x_{N-1}
\end{bmatrix}^\top$
into
an
$N$-point
uncorrelated signal
$\mathbf{y} =
\begin{bmatrix}
	y_0 & y_1 & \ldots & y_{N-1}
\end{bmatrix}^\top$
according
to
\begin{align}
	\mathbf{y} = \mathbf{K} \cdot \mathbf{x},
\end{align}
where
$\mathbf{K}$ is the KLT matrix~\cite{karhunen1947under,loeve1948functions}.
If
$\mathbf{x}$
is
a first-order Markovian signal,
then it was shown in~\cite{jain1976fast} that
	the $(i,j)$th entry of the KLT matrix for a
	given value of the correlation coefficient $\rho \in [0,1]$ is~\cite{britanak2010discrete}:
	\begin{equation}\label{eq:u}
		k_{i,j} = \sqrt{\frac{2}{N+ \lambda_i}} \sin \left[\omega_i \left(i - \frac{N-1}{2}\right)+ \frac{(j+1)\pi}{2}\right],
	\end{equation}
	where
	$i,j = 0,1,\ldots,N-1$,
	$
	\lambda_i = \frac{1-\rho^2}{1+\rho^2 -2\rho \cos \omega_i}
	$,
	and
	$\omega_1, \omega_2, \ldots, \omega_{N}$ are the solutions to
	$
	\tan N \omega = \frac{-(1-\rho^2)\sin \omega}{(1+\rho^2)\cos \omega - 2\rho}
	$~\cite{britanak2010discrete}.

	\subsection{Approximation Theory}\label{ss:approxtheory}

	Generally, the approximation $\widehat{\mathbf{K}}$ is based on a low-complexity matrix $\mathbf{T}$, such that $\widehat{\mathbf{K}} = \mathbf{S} \cdot \mathbf{T}$~\cite{cintra2014low,oliveira2015discrete, tablada2017dct, oliveira2019low}, and
	\begin{equation}
		\label{eq:diag}
		\mathbf{S} =
		\begin{cases}
			\sqrt{(\mathbf{T}\cdot\mathbf{T}^\top)^{-1}}, &  \text{if $\mathbf{T}$ is orthogonal,}\\
			\sqrt{[\operatorname{diag}(\mathbf{T}\cdot\mathbf{T}^\top)]^{-1}}, &  \text{if $\mathbf{T}$ is non-orthogonal,}
		\end{cases}
	\end{equation}
	where $\operatorname{diag}(\cdot)$ is the diagonal matrix generated by its arguments.

	Thus, we focus our search on the matrix $\mathbf{T}$.
	The low-complexity matrix $\mathbf{T}$ can be obtained
	by restricting its elements over
	sets whose entries possess very low multiplicative complexity, such as
	$\{0, \pm 1, \pm 2 \}$,  $\{0, \pm 1/2, \pm 1, \pm2\}$, $\{0, \pm1, \pm2, \pm3\}$, among others.
	As a matter of fact, multiplications by powers of two require
	only bit-shifting operations
	and a multiplication by $3$
	can be implemented by means of one addition and one bit-shifting operation.
	A possible way of restricting the entries of matrix $\mathbf{T}$ is applying integer functions to the elements of the exact transform, as shown in~\cite{haweel2001new,cintra2011dct,cintra2014low}.
	Common integer functions employed to derive new transform approximations are the floor, ceiling,  truncation (round towards zero), and round-away-from-zero functions, defined respectively as:
	\begin{align}\label{eq:integerfunctions}
		\begin{split}
			& \operatorname{floor}(x) = \lfloor x \rfloor = \operatorname{max}\{m \in \mathbb{Z} \mid m \leq x\},\\
			& \operatorname{ceil}(x)  = \lceil x \rceil = \operatorname{min}  \{n \in \mathbb{Z} \mid n \geq x \} \\
			& \operatorname{trunc}(x) = \operatorname{sign}(x)\cdot \lfloor\mid x \mid\rfloor,\\
			& \operatorname{round}_\text{AFZ}(x) = \operatorname{sign}(x)\cdot \lceil\mid x\mid \rceil,
		\end{split}
	\end{align}
	where $|\cdot|$ is the absolute value of its argument.

	\subsection{Assessment Metrics}\label{ss:measures}
	The performance measures usually employed for assessing approximate transforms can be categorized in two types:
	(i)~coding measures, such as the coding gain~\cite{katto1992short} and transform efficiency~\cite{nikara2001unified}, which measure the power of energy decorrelation and compaction; and (ii)~proximity measures relative to the exact transform, such as mean square error~\cite{britanak2010discrete}
	and total energy error~\cite{cintra2011dct}, which quantify similarities between the approximate matrices and the exact transform in a Euclidean distance sense. Such figures of merit are presented next.
	\subsubsection{Unified Coding Gain}
	The unified coding gain measures the energy compaction  capacity  and is given by~\cite{katto1992short}:
	\begin{equation}\label{eq:cg}
		{\textrm{C}_g} ({\widehat{\mathbf{K}}}) =
		10 \cdot
		\log_{10}
		\Biggl\{
		\prod_{k=1}^N
		\frac{1}{\sqrt[N]{A_k \cdot B_k}}
		\Biggr\}
		,
	\end{equation}
	where
	$A_k = \operatorname{su} \left\{ (\mathbf{h}_k^\top \cdot \mathbf{h}_k)\odot \mathbf{R}_{\mathbf{x}}
	\right\}$,
	$\mathbf{h}_k$ is the $k$th row vector from $\widehat{\mathbf{K}}$,
	function  $\operatorname{su}(\cdot)$ gives the sum of the elements of its matrix argument,
	$\odot$ is the Hadamard matrix product~\cite{seber2008matrix},
	$\mathbf{R}_{\mathbf{x}}$ is the autocorrelation matrix from a first-order Markovian signal,
	$B_k = \| \mathbf{g}_k \|_\mathrm{F}^2$,
	$\mathbf{g}_k$ is the $k$th row vector from~$\widehat{\mathbf{K}}^{-1}$, and $\|\cdot \|_\mathrm{F}$ is the Frobenius norm~\cite{seber2008matrix}.
	\subsubsection{Transform Efficiency}
	The transform efficiency is given by~\cite{nikara2001unified}:
	\begin{equation}\label{eq:eta}
		\eta({\widehat{\mathbf{K}}}) = 100 \frac{ \sum_{i = 1}^{N} \vert r_{i,i}\vert}{ \sum_{i = 1}^{N}  \sum_{j = 1}^{N} \vert r_{i,j} \vert},
	\end{equation}
	where $r_{i,j}$ is the $(i,j)$th element from
	$\widehat{\mathbf{K}}\cdot
	\mathbf{R}_{\mathbf{x}}
	\cdot \widehat{\mathbf{K}}^\top$.

	\subsubsection{Mean square Error}
	The mean square error between the exact and approximate transforms is defined as~\cite{britanak2010discrete}:
	\begin{equation}\label{eq:MSE}
		{\textrm{MSE}}(\mathbf{K},\widehat{\mathbf{K}}) = \frac{1}{N}\cdot
		\operatorname{tr}
		\left\{
		(\mathbf{K} - {\widehat{\mathbf{K}}})\cdot \mathbf{R}_{\mathbf{x}} \cdot (\mathbf{K} - {\widehat{\mathbf{K}}})^\top
		\right\}
		,
	\end{equation}
	where $\operatorname{tr}(\cdot)$ is the trace function~\cite{harville1997trace}.
	\subsubsection{Total Error Energy}
	The total error energy measures the similarity between the approximate and the exact transform matrix, according to~\cite {cintra2011dct}:
	\begin{equation}\label{eq:ep}
		\epsilon(\mathbf{K},\widehat{\mathbf{K}}) = \pi \cdot \|\mathbf{K} - {\widehat{\mathbf{K}}}\|_\mathrm{F}^2.
	\end{equation}

	\section{Optimal Proposed Transforms}\label{S:ProposedTransform}
	\subsection{Search Space}\label{ss:search}

	For the computational search,
	we set the elements of the matrices to be in the set of low-complexity entries $\mathcal{C} =  \{0, \pm 1, \pm 2, \pm 3 \} $ since the multiplication by this elements require only additions and bit-shifting operations.
	For the blocklength we considered  $N = 8$,
	due to its importance in image compression.
	Thus,
	we have $7^8 = 5764801$ candidate matrices to be considered in the optimization problem for each value of $\rho$ of the KLT matrix.

	The transform search space can be formally defined as follows.
	Let
	\begin{equation}
		\mathbf{\widehat{K}} = \sqrt{[\operatorname{diag}(\mathbf{T}\cdot \mathbf{T}^\top)]^{-1}} \cdot \mathbf{T},
	\end{equation}
	where $\mathbf{T} \in \mathcal{M}_\mathcal{C}(8)$  and $\mathcal{M}_\mathcal{C}(8)$ is the $8 \times 8$ matrix space with elements in the set $\mathcal{C} = \{0, \pm 1, \pm 2, \pm 3 \}$. We propose to search a subset of $\mathcal{M}_\mathcal{C}(8)$:
	\begin{equation}
		\mathcal{E}_\alpha = \{\mathbf{T} \in \mathcal{M}_\mathcal{C}(8):  \mathbf{T} = \operatorname{int}(\alpha \cdot \mathbf{K}^{(\rho)})  \},
	\end{equation}
	where  $\operatorname{int} \in \{\operatorname{floor}, \operatorname{ceil}, \operatorname{trunc}, \operatorname{round}_\text{AFZ}\}$,
	and $\alpha$ is the expansion factor~\cite{plonka2004global,britanak2010discrete,cintra2014low}.
	The ranges of $\alpha$ must satisfy the inequality $0 \leq \operatorname{int}(\alpha \cdot \gamma) \leq 3$,
	where $\gamma$ is the absolute value of the largest element of the matrix $\mathbf{K}^{(\rho)}$.
	Considering the integer functions $\operatorname{floor}$, $\operatorname{ceil}$, $\operatorname{trunc}$, and $\operatorname{round}_\text{AFZ}$, the ranges of $\alpha$ ($\mathcal{A}$)
	are given, respectively, by: $(1/\gamma, 4/\gamma)$, $(0, 3/\gamma)$, $(1/\gamma, 4/\gamma)$, and $(0, 3/\gamma)$.
	Therefore, the search space is:
	\begin{equation}
		\mathcal{E} = \bigcup_{\alpha \in \mathcal{A}} \mathcal{E}_\alpha.
	\end{equation}

	\subsection{Objective Function}\label{ss:objective}
	In order to search for the optimal transforms according to the considered metrics, the following optimization problem was proposed:
	\begin{equation}\label{eq:opt}
		\mathbf{\widehat{K}}^* = \operatornamewithlimits{arg \ opt}_{\widehat{\mathbf{K}}} f(\mathbf{\widehat{K}}),
	\end{equation}
	where $\widehat{\mathbf{K}}$ is a candidate matrix  for solving the problem, and
	\begin{equation}
		f \in \{ {\textrm{C}_g} (\cdot), \eta(\cdot), {\textrm{MSE}}(\mathbf{K}^{(\rho)},\cdot), \epsilon(\mathbf{K}^{(\rho)},\cdot)\},
	\end{equation}
	which are the figures of merit to be optimized.
	For ${\textrm{C}_g(\cdot)}$ and $\eta(\cdot)$, the optimization problem is of the maximization type; whereas, for the ${\textrm{MSE}(\cdot)}$ and $\epsilon(\cdot)$, it is of the minimization type.

	\subsection{Methodology}
	Once the optimization problem, restrictions, search space, and the objective function are established,
	we exhaustively compute \eqref{eq:opt} for the specific values of $\alpha$ within the intervals defined by each integer function, with steps of $10^{-2}$. To compute the exact KLT matrix, $\mathbf{K}^{(\rho)}$, we considered $0 < \rho < 1$ with steps of $10^{-1}$.

	This search results in $144$ matrices. Here we are considering the discussed four figures of merit and evaluating each one separately, so we can have up to four optimal transforms for each fixed interval of $\rho$ and integer function.
	Among the $144$ obtained matrices, it is expected that several of them show similar performance.
	Therefore, we aim now at refining the set of $144$ matrices so we could identify a reduced set of matrices that are
	representative over the range $\rho \in (0,1)$.

	In this sense,
	we propose a two-step procedure. In the first step, we only consider, among the 144 transforms, those that obtained the best performance, for the values of $\rho \in (0,1)$ with steps of $10^{-1}$, according to each figure of merit. This procedure caused a reduction of $86.11\%$
	in the number of transforms. Table~\ref{t:measurestransf} presents the
	$20$ transforms that exhibits the best performance between all obtained transforms.
	The similarity measurements were obtained considering the exact KLT for the value of $\rho$ of the upper limit of each interval from which the approximate was derived, i.e., $\widehat{\mathbf{K}}_1$ were compared to the exact KLT for $\rho = 0.1$.

	\begin{table}[hbt]
		\caption{Coding and similarity measures from the obtained optimal transforms}
		\label{t:measurestransf}
		\centering
		\begin{tabular}{lcccccc}
			\toprule
			\addlinespace[0.8ex]
			Transform
			& $\rho$ interval & ${\textrm{C}_g} (\widehat{\mathbf{K}})$ & $\eta(\widehat{\mathbf{K}})$ & $\epsilon(\mathbf{K}^{(\rho)}, \widehat{\mathbf{K}})$ & ${\textrm{MSE}}(\mathbf{K}^{(\rho)}, \widehat{\mathbf{K}})$\\
			\addlinespace[0.8ex]
			\midrule
			\addlinespace[0.8ex]
			$\widehat{\mathbf{K}}_1$
			& $(0,0.1]$	&	$	0.0308	$	&	$	93.4298	$	&	$	1.5331	$	&	$	0.0608	$	\\
			$\widehat{\mathbf{K}}_2$
			& $(0,0.1]$	&	$	0.1325	$	&	$	79.5971	$	&	$	0.3173	$	&	$	0.0128	$	\\
			$\widehat{\mathbf{K}}_3$
			& $(0,0.1]$	&	$	0.0588	$	&	$	88.3104	$	&	$	0.093	$	&	$	0.0036$	\\
			$\widehat{\mathbf{K}}_4$
			& $(0.1,0.2]$	&	$	0.1754	$	&	$	83.7756	$	&	$	0.2265	$	&	$	0.0094	$	\\
			$\widehat{\mathbf{K}}_{5}$
			& $(0.2,0.3]$	&	$	0.3461	$	&	$	80.8238	$	&	$	0.2999	$	&	$	0.0132	$	\\
			$\widehat{\mathbf{K}}_{6}$
			& $(0.3,0.4]$	&	$	0.6725	$	&	$	83.0728	$	&	$	0.3104	$	&	$	0.0095	$	\\
			$\widehat{\mathbf{K}}_{7}$
			& $(0.3,0.4]$	&	$	0.7618	$	&	$	63.712	$	&	$	2.1348	$	&	$	0.0785	$	\\
			$\widehat{\mathbf{K}}_{8}$
			& $(0.3,0.4]$	&	$	0.6532	$	&	$	83.3729	$	&	$	0.2823	$	&	$	0.0115	$	\\
			$\widehat{\mathbf{K}}_{9}$
			& $(0.4,0.5]$	&	$	1.1063	$	&	$	87.2737	$	&	$	0.3487	$	&	$	0.0094	$	\\
			$\widehat{\mathbf{K}}_{10}$
			& $(0.4,0.5]$	&	$	1.153	$	&	$	77.5984	$	&	$	0.6439	$	&	$	0.0163	$	\\
			$\widehat{\mathbf{K}}_{11}$
			& $(0.5,0.6]$ &	$	1.7572	$	&	$	82.7462	$	&	$	0.9273	$	&	$	0.0197	$	\\
			$\widehat{\mathbf{K}}_{12}$
			& $(0.5,0.6]$&	$	1.6743	$	&	$	86.4929	$	&	$	0.275	$	&	$	0.0089	$	\\
			$\widehat{\mathbf{K}}_{13}$
			& $(0.6,0.7]$	&	$	2.5736	$	&	$	84.7636	$	&	$	0.7505	$	&	$	0.0153	$	\\
			$\widehat{\mathbf{K}}_{14}$
			& $(0.6,0.7]$	&	$	2.5308	$	&	$	89.7579	$	&	$	0.2299	$	&	$	0.0065	$	\\
			$\widehat{\mathbf{K}}_{15}$
			& $(0.7,0.8]$&	$	3.8534	$	&	$	84.1782	$	&	$	0.6043	$	&	$	0.0087	$	\\
			$\widehat{\mathbf{K}}_{16}$
			& $(0.7,0.8]$&	$	3.8484	$	&	$	87.7103	$	&	$	0.2418	$	&	$	0.0043	$	\\
			$\widehat{\mathbf{K}}_{17}$
			& $(0.7,0.8]$	&	$	3.8146	$	&	$	86.6308	$	&	$	0.1884	$	&	$	0.0049	$	\\
			$\widehat{\mathbf{K}}_{18}$
			& $(0.8,1)$	&	$	6.2462	$	&	$	88.1734	$	&	$	0.6746	$	&	$	0.0102	$	\\
			$\widehat{\mathbf{K}}_{19}$
			& $(0.8,1)$	&	$	6.1727	$	&	$	85.8301	$	&	$	0.1948	$	&	$	0.0055	$	\\
			$\widehat{\mathbf{K}}_{20}$
			& $(0.8,1)$&	$	6.2335	$	&	$	86.827	$	&	$	0.4439	$	&	$	0.005	$	\\
			\addlinespace[0.8ex]
			\bottomrule
		\end{tabular}
	\end{table}

	In the second stage of the refinement, we aim to group intervals of $\rho$ in which the transforms exhibit similar performance according to the unified coding gain, since this metric presents information about the coding capacity of the orthogonal transformation for applications of data compression.
	These groups  can be obtained according to a clustering procedure, such as
	the $k$-means~\cite{hartigan1979algorithm}. Using a clustering method can result in a reduced number of groups in which only one matrix can be chosen as representative of the group.
	Fig.~\ref{f:intervalosrho} represents graphically the idea: we intend to find transforms $\mathbf{\widehat{K}}_*$ that represent the transforms for some interval of $\rho$.
	\begin{figure}[h!]
		\centering
		\includegraphics[width=10cm,height=5cm]{intervalos.eps}
		\caption{%
Example of $\rho$ intervals in groups.}\label{f:intervalosrho}
	\end{figure}

The selection of the k-means clustering algorithm was based on its simplicity and widespread use in unsupervised learning. This method efficiently addresses clustering problems, making it a suitable choice for this study~\cite{flach2012machine}. The pseudocode for the $k$-means clustering method is presented in Algorithm~\ref{alg:k-means}.

\begin{algorithm}
\caption{%
Pseudocode for the $k$-means Clustering Method}\label{alg:k-means}
\begin{algorithmic}
	\Require Data: $D\subseteq\mathbb{R}^d$;  Number of clusters: $C\in\mathbb{N}$.
	\Ensure  $C$ clusters means: $\mu_1,\ldots,\mu_c \in \mathbb{R}^d$
	\State Randomly initialize the $C$ vectors $\mu_1,\ldots,\mu_c \in \mathbb{R}^d$
	\While{There are no more changes to $\mu_1,\ldots,\mu_c$}
	\State{Attribute $x\in D$ for $\operatornamewithlimits{arg \ min}_{j} \operatorname{Dis}(x,\mu_j)$;}
	\For{$j=1$ to $C$}
	\State $D_j \leftarrow \{x \in D \mid x \quad\text{attributed to cluster}\quad C\}$;
	\State $\mu_j = \frac{1}{\mid D_j\mid}\sum_{x\in D_j} x$;
	\EndFor
	\EndWhile
\end{algorithmic}
\end{algorithm}

	Hence, applying the $k$-means clustering method considering the values of the unified coding gain of each transform, we obtained two distinct groups, $C_1$ and $C_2$. The first group presented transforms for the values of $\rho$ ranging from $(0,0.7]$ and the second group transforms for values of $\rho \in (0.7,1)$.
	In each group, we considered the transforms which presented the best values of the discussed figures of merit. The optimal transforms chosen are presented in the next section.
It is important to highlight that the methodology discussed here represents a novel approach within the literature concerning KLT, with potential applicability across various discrete transforms and block-lengths $N$.

	\section{Proposed Approximate KLT and Evaluation}\label{S:results}
	For the group $C_1$, which represents the transforms obtained for values of $\rho \in(0,0.7]$,  the optimal transforms are $\widehat{\mathbf{K}}_1$, $\widehat{\mathbf{K}}_3$, and $\widehat{\mathbf{K}}_{13}$. For $C_2$, the transforms $\widehat{\mathbf{K}}_{16}$, $\widehat{\mathbf{K}}_{17}$, and $\widehat{\mathbf{K}}_{18}$ were the ones which perform better considering values of $\rho \in (0.7,1)$. Table~\ref{t:transforms} presents the low-complexity matrices that generate the respective transforms.

	\begin{table}[h!]
		\caption{Low-complexity matrices of the KLT Approximations}
		\label{t:transforms}
		\centering
		\scriptsize
		\begin{tabular}{lccccc}
			\toprule
			Transform  & Matrix  \\
			\midrule
			$\mathbf{T}_1$
			&
			$\left[ \begin{matrix}
				0 & 1 & 1 & 1 & 1 & 1 & 1 & 0 \\
				1 & 1 & 1 & 0 & 0 & -1 & -1 & -1 \\
				1 & 1 & 0 & -1 & -1 & 0 & 1 & 1 \\
				1 & 0 & -1 & -1 & 1 & 1 & 0 & -1 \\
				1 & 0 & -1 & 1 & 1 & -1 & 0 & 1 \\
				1 & -1 & 0 & 1 & -1 & 0 & 1 & -1 \\
				1 & -1 & 1 & 0 & 0 & 1 & -1 & 1 \\
				0 & -1 & 1 & -1 & 1 & -1 & 1 & 0 \\
			\end{matrix} \right]$
			\\
			$\mathbf{T}_3$
			&
			$\left[ \begin{matrix}
				1 & 2 & 3 & 3 & 3 & 3 & 2 & 1 \\
				2 & 3 & 3 & 1 & -1 & -3 & -3 & -2 \\
				3 & 3 & 0 & -3 & -3 & 0 & 3 & 3 \\
				3 & 1 & -3 & -2 & 2 & 3 & -1 & -3 \\
				3 & -1 & -3 & 2 & 2 & -3 & -1 & 3 \\
				3 & -3 & 0 & 3 & -3 & 0 & 3 & -3 \\
				2 & -3 & 3 & -1 & -1 & 3 & -3 & 2 \\
				1 & -2 & 3 & -3 & 3 & -3 & 2 & -1 \\
			\end{matrix} \right]$
			\\
			$\mathbf{T}_{13}$
			&
			$\left[ \begin{matrix}
				1 & 1 & 1 & 2 & 2 & 1 & 1 & 1 \\
				2 & 2 & 1 & 0 & 0 & -1 & -2 & -2 \\
				2 & 1 & 0 & -2 & -2 & 0 & 1 & 2 \\
				2 & 0 & -2 & -1 & 1 & 2 & 0 & -2 \\
				1 & -1 & -1 & 1 & 1 & -1 & -1 & 1 \\
				1 & -2 & 0 & 2 & -2 & 0 & 2 & -1 \\
				1 & -2 & 2 & -1 & -1 & 2 & -2 & 1 \\
				0 & -1 & 2 & -2 & 2 & -2 & 1 & 0 \\
			\end{matrix} \right]$
				\\
				$\mathbf{T}_{16}$
				&
				$\left[ \begin{matrix}
					2 & 2 & 2 & 2 & 2 & 2 & 2 & 2 \\
					3 & 3 & 2 & 1 & -1 & -2 & -3 & -3 \\
					3 & 2 & -1 & -3 & -3 & -1 & 2 & 3 \\
					3 & 0 & -3 & -2 & 2 & 3 & 0 & -3 \\
					2 & -2 & -2 & 2 & 2 & -2 & -2 & 2 \\
					2 & -3 & 1 & 2 & -2 & -1 & 3 & -2 \\
					1 & -3 & 3 & -1 & -1 & 3 & -3 & 1 \\
					1 & -2 & 3 & -3 & 3 & -3 & 2 & -1 \\
				\end{matrix} \right]$
				\\
				$\mathbf{T}_{17}$
				&
				$\left[ \begin{matrix}
					2 & 2 & 2 & 2 & 2 & 2 & 2 & 2 \\
					3 & 3 & 2 & 1 & -1 & -2 & -3 & -3 \\
					3 & 2 & -1 & -3 & -3 & -1 & 2 & 3 \\
					3 & 0 & -3 & -2 & 2 & 3 & 0 & -3 \\
					2 & -2 & -2 & 2 & 2 & -2 & -2 & 2 \\
					2 & -3 & 1 & 3 & -3 & -1 & 3 & -2 \\
					1 & -3 & 3 & -1 & -1 & 3 & -3 & 1 \\
					1 & -2 & 3 & -3 & 3 & -3 & 2 & -1 \\
				\end{matrix} \right]$
				\\
				$\mathbf{T}_{18}$
				&
				$\left[ \begin{matrix}
					1 & 1 & 1 & 2 & 2 & 1 & 1 & 1 \\
					2 & 2 & 1 & 0 & 0 & -1 & -2 & -2 \\
					2 & 1 & -1 & -2 & -2 & -1 & 1 & 2 \\
					2 & 0 & -2 & -1 & 1 & 2 & 0 & -2 \\
					1 & -1 & -1 & 1 & 1 & -1 & -1 & 1 \\
					1 & -2 & 0 & 2 & -2 & 0 & 2 & -1 \\
					1 & -2 & 2 & -1 & -1 & 2 & -2 & 1 \\
					0 & -1 & 2 & -2 & 2 & -2 & 1 & 0 \\
				\end{matrix} \right]$
				\\
				\bottomrule
			\end{tabular}
		\end{table}

		Table~\ref{t:measurescomp8} presents the coding and similarity measurements of the exact KLT for a given value of $\rho$ and the proposed approximate transforms.
		We divided Table~\ref{t:measurescomp8} into the two groups, $C_1$ and $C_2$,
		and compared the approximate transforms from group $C_1$ to the exact KLT for $\rho = 0.2$ ($\mathbf{K}_8^{(0.2)}$), and the approximate transforms from group $C_2$ to the exact KLT for $\rho = 0.8$ ($\mathbf{K}_8^{(0.8)}$).
We can note that the proposed approximations exhibit comparable performance to the exact KLT for a specific value of $\rho$, further confirming the viability of using the approximation as a substitute for the exact transform.

		\begin{table}[hbt] \caption{Comparison of coding and similarity measures between the exact KLT and the proposed approximate transforms}
			\label{t:measurescomp8}
			\centering
			\begin{tabular}{lccccc}
				\toprule
				\addlinespace[0.8ex]
				Transform & ${{\textrm{C}_g}}(\widehat{\mathbf{K}})$ & $\eta(\widehat{\mathbf{K}})$ & $\epsilon(\mathbf{K}^{(\rho)}, \widehat{\mathbf{K}})$ & ${\textrm{MSE}}(\mathbf{K}^{(\rho)}, \widehat{\mathbf{K}})$\\
				\midrule
				$\mathbf{K}_8^{(0.2)}$ &	$	0.1551 $	&	$	100	$	&	$	0	$	&	$	0	$	\\
								$\mathbf{\widehat{K}}_1$ 	&	$	0.0308	$	&	$	93.4298	$	&	$	1.5331	$	&	$	0.0608	$	\\
				$\mathbf{\widehat{K}}_3$ 	&	$	0.0588	$	&	$	88.3104	$	&	$	0.093	$	&	$	0.0036	$	\\
				$\mathbf{\widehat{K}}_{13}$ 	&	$	2.5736$	&	$	84.7636	$	&	$	0.7505	$	&	$	0.0153	$\\
				\addlinespace[1ex]
				\hdashline
				\addlinespace[1ex]
				$\mathbf{K}_8^{(0.8)}$ &	$	 3.8824
				$	&	$	100	$	&	$	0	$	&	$	0	$	\\
				$\mathbf{\widehat{K}}_{16}$ & 	$	3.8484	$	&	$	87.7103	$	&	$	0.2418	$	&	$0.0043	$\\
				$\mathbf{\widehat{K}}_{17}$	& $	3.8146	$	&	$	86.6308	$	&	$	0.1884	$	&	$	0.0049	$		\\
				$\mathbf{\widehat{K}}_{18}$	&	$	6.2462	$	&	$	88.1734	$	&	$	0.6746	$	&	$	0.0102	$	\\
				\bottomrule
			\end{tabular}
		\end{table}

		\section{Fast Algorithms and Computational Complexity}\label{S:fastalgo}

		\subsection{Proposed Fast Algorithms}
		By factoring the matrices of the proposed optimal transforms, $\mathbf{T}_1$, $\mathbf{T}_3$, $\mathbf{T}_{13}$, $\mathbf{T}_{16}$, $\mathbf{T}_{17}$, and $\mathbf{T}_{18}$ into sparse matrices, considering butterfly-based structures~\cite{blahut2010fast}, we obtain the following decomposition:

		\begin{align}
			\mathbf{T}_i =& \mathbf{P}\cdot \mathbf{M}_i \cdot \mathbf{A}_1, \qquad\qquad i = 1,3,13,
			\label{eq:f1}\\
			\mathbf{T}_j =& \mathbf{P} \cdot \mathbf{M}_j \cdot \mathbf{A}_2^\prime \cdot \mathbf{A}_1, \qquad j = 16,17, \label{eq:f2}\\
			\mathbf{T}_{18} =& \mathbf{P} \cdot \mathbf{M}_{18} \cdot \mathbf{A}_2^{\prime\prime} \cdot \mathbf{A}_1,  \label{eq:f3}
		\end{align}
		where $\mathbf{P}$ is a permutation matrix,  $\mathbf{A}_1$, $\mathbf{A}_2^\prime$, and  $\mathbf{A}_2^{\prime\prime}$ are additive matrices, and $\mathbf{M}$ is a multiplicative matrix.
		For the factorization of  $\mathbf{T}_1$, $\mathbf{T}_3$, and $\mathbf{T}_{13}$, we have:
		\begin{equation}\label{eq:P} %
			\mathbf{P} =
			\left[
			\begin{rsmallmatrix}
				1 & 0 & 0 & 0 & 0 & 0 & 0 & 0 \\
				0 & 0 & 0 & 0 & 1 & 0 & 0 & 0 \\
				0 & 1 & 0 & 0 & 0 & 0 & 0 & 0 \\
				0 & 0 & 0 & 0 & 0 & 1 & 0 & 0 \\
				0 & 0 & 1 & 0 & 0 & 0 & 0 & 0 \\
				0 & 0 & 0 & 0 & 0 & 0 & 1 & 0 \\
				0 & 0 & 0 & 1 & 0 & 0 & 0 & 0 \\
				0 & 0 & 0 & 0 & 0 & 0 & 0 & 1 \\
			\end{rsmallmatrix}
			\right], \quad
			\mathbf{A}_1 =
			\left[
			\begin{rsmallmatrix}
				1 & 0 & 0 & 0 & 0 & 0 & 0 & 1 \\
				0 & 1 & 0 & 0 & 0 & 0 & 1 & 0 \\
				0 & 0 & 1 & 0 & 0 & 1 & 0 & 0 \\
				0 & 0 & 0 & 1 & 1 & 0 & 0 & 0 \\
				0 & 0 & 0 & 1 & -1 & 0 & 0 & 0 \\
				0 & 0 & 1 & 0 & 0 & -1 & 0 & 0 \\
				0 & 1 & 0 & 0 & 0 & 0 & -1 & 0 \\
				1 & 0 & 0 & 0 & 0 & 0 & 0 & -1 \\
			\end{rsmallmatrix}
			\right].
		\end{equation}
		The multiplicative matrix $\mathbf{M}$ can be written as:
		\begin{equation}\label{eq:Ki} %
			\mathbf{M}=
			\left[
			\begin{matrix}
				\mathbf{M}_1 & \\
				& \mathbf{M}_2 \\
			\end{matrix}
			\right],
		\end{equation}
		where
		\begin{equation}\label{eq:matrixA1} %
			\mathbf{M}_1 = \mathbf{M}_2 =
			\left[
			\begin{matrix}
				m_0 & m_1 & m_2 & m_3 \\
				m_4& m_5 & m_6 & m_7 \\
				m_8 & m_9 & m_{10} & m_{11} \\
				m_{12} & m_{13} & m_{14} & m_{15} \\
			\end{matrix}
			\right],
		\end{equation}
		and the constants $m_k$, $k = 0,1, \ldots, 15$ depend on the choice of the matrix $\mathbf{T}$ and are presented in the Table~\ref{t:emes}.
		For the factorization of $\mathbf{T}_{16}$, $\mathbf{T}_{17}$, and for $\mathbf{T}_{18}$ we have:
		\begin{equation}\label{eq:matrixA2} %
			\mathbf{A}_2^\prime =
			\left[
			\begin{rsmallmatrix}
				1 & 0 & 0 & 1 & 0 & 0 & 0 & 0 \\
				0 & 1 & 0 & 0 & 0 & 0 & 0 & 0 \\
				0 & 0 & 1 & 0 & 0 & 0 & 0 & 0 \\
				1 & 0 & 0 & -1 & 0 & 0 & 0 & 0 \\
				0 & 0 & 0 & 0 & 1 & 0 & 0 & 0 \\
				0 & 0 & 0 & 0 & 0 & 1 & 0 & 0 \\
				0 & 0 & 0 & 0 & 0 & 0 & 1 & 0 \\
				0 & 0 & 0 & 0 & 0 & 0 & 0 & 1 \\
			\end{rsmallmatrix}
			\right], \quad
			\mathbf{A}_2^{\prime\prime} =
			\left[
			\begin{rsmallmatrix}
				1 & 0 & 0 & 0 & 0 & 0 & 0 & 0 \\
				0 & 1 & 1 & 0 & 0 & 0 & 0 & 0 \\
				0 & 1 & -1 & 0 & 0 & 0 & 0 & 0 \\
				0 & 0 & 0 & 1 & 0 & 0 & 0 & 0 \\
				0 & 0 & 0 & 0 & 1 & 0 & 0 & 0 \\
				0 & 0 & 0 & 0 & 0 & 1 & 0 & 0 \\
				0 & 0 & 0 & 0 & 0 & 0 & 1 & 0 \\
				0 & 0 & 0 & 0 & 0 & 0 & 0 & 1 \\
			\end{rsmallmatrix}
			\right].
		\end{equation}

		\begin{table*}[hbt]
			\caption{Constants required for the fast algorithm for blocks $\mathbf{M}_1$ and $\mathbf{M}_2$}
			\label{t:emes}
			\tiny
			\centering
			\begin{tabular}{crrrrrrrrrrrrrr}
				\toprule
				\multicolumn{1}{l}{\multirow{2}{*}{Constants}} & \multicolumn{2}{c}{$\mathbf{T}_1$ } & \multicolumn{2}{c}{$\mathbf{T}_3$ } & \multicolumn{2}{c}{$\mathbf{T}_{13}$} & \multicolumn{2}{c}{$\mathbf{T}_{16}$} & \multicolumn{2}{c}{$\mathbf{T}_{17}$} & \multicolumn{2}{c}{$\mathbf{T}_{18}$ }       \\
				\multicolumn{1}{l}{}          & $\mathbf{M}_1$       & $\mathbf{M}_2$        & $\mathbf{M}_1$         & $\mathbf{M}_2$        & $\mathbf{M}_1$       & $\mathbf{M}_2$          & $\mathbf{M}_1$          & $\mathbf{M}_2$          & $\mathbf{M}_1$        & $\mathbf{M}_2$          & $\mathbf{M}_1$         & $\mathbf{M}_2$         \\  \midrule
				$m_0$	&	$	0	$	&	$	0	$	&	$	1	$	&	$	1	$	&	$	1	$	&	$	0	$	&	$	2	$	&	$	1	$	&	$	2	$	&	$	1	$	&	$	1	$	&	$	0	$	\\
				$m_1$	&	$	1	$	&	$	1	$	&	$	2	$	&	$	3	$	&	$	1	$	&	$	1	$	&	$	2	$	&	$	2	$	&	$	2	$	&	$	2	$	&	$	1	$	&	$	1	$	\\
				$m_2$	&	$	1	$	&	$	1	$	&	$	3	$	&	$	3	$	&	$	1	$	&	$	2	$	&	$	2	$	&	$	3	$	&	$	2	$	&	$	3	$	&	$	0	$	&	$	2	$	\\
				$m_3$	&	$	1	$	&	$	1	$	&	$	3	$	&	$	2	$	&	$	2	$	&	$	2	$	&	$	0	$	&	$	3	$	&	$	0	$	&	$	3	$	&	$	2	$	&	$	2	$	\\
				$m_4$	&	$	1	$	&	$	-1	$	&	$	3	$	&	$	-2	$	&	$	2	$	&	$	-1	$	&	$	0	$	&	$	-2	$	&	$	0	$	&	$	-2	$	&	$	2	$	&	$	-1	$	\\
				$m_5$	&	$	1	$	&	$	-1	$	&	$	3	$	&	$	-3	$	&	$	1	$	&	$	-2	$	&	$	2	$	&	$	-3	$	&	$	2	$	&	$	-3	$	&	$	0	$	&	$	-2	$	\\
				$m_6$	&	$	0	$	&	$	0	$	&	$	0	$	&	$	1	$	&	$	0	$	&	$	0	$	&	$	-1	$	&	$	0	$	&	$	-1	$	&	$	0	$	&	$	1	$	&	$	0	$	\\
				$m_7$	&	$	-1	$	&	$	1	$	&	$	-3	$	&	$	3	$	&	$	-2	$	&	$	2	$	&	$3	$	&	$	3	$	&	$	3	$	&	$	3	$	&	$-2	$	&	$	2	$	\\
				$m_8$	&	$	1	$	&	$	1	$	&	$	3	$	&	$	3	$	&	$	1	$	&	$	2	$	&	$	2	$	&	$	2	$	&	$	2	$	&	$	3	$	&	$	1	$	&	$	2	$	\\
				$m_9$	&	$	0	$	&	$	0	$	&	$	-1	$	&	$	0	$	&	$	-1	$	&	$	0	$	&	$	-2	$	&	$	1	$	&	$	-2	$	&	$	1	$	&	$	-1	$	&	$	0	$	\\
				$m_{10}$	&	$	-1	$	&	$	-1	$	&	$	-3	$	&	$	-3	$	&	$	-1	$	&	$	-2	$	&	$	-2	$	&	$	-3	$	&	$	-2	$	&	$	-3	$	&	$	0	$	&	$	-2	$	\\
				$m_{11}$	&	$	1	$	&	$	1	$	&	$	2	$	&	$	3	$	&	$	1	$	&	$	1	$	&	$	0	$	&	$	2	$	&	$	0	$	&	$	2	$	&	$	1	$	&	$	1	$	\\
				$m_{12}$	&	$	1	$	&	$	-1	$	&	$	2	$	&	$	-3	$	&	$	1	$	&	$	-2	$	&	$	0	$	&	$	-3	$	&	$	0	$	&	$	-3	$	&	$	1	$	&	$	-2	$	\\
				$m_{13}$	&	$	-1	$	&	$	1	$	&	$	-3	$	&	$	3	$	&	$	-2	$	&	$	2	$	&	$	-3	$	&	$	3	$	&	$	-3	$	&	$	3	$	&	$	0	$	&	$	2	$	\\
				$m_{14}$	&	$	1	$	&	$	-1	$	&	$	3	$	&	$	-2	$	&	$	2	$	&	$	-1	$	&	$	3	$	&	$	-2	$	&	$	3	$	&	$	-2	$	&	$	-2	$	&	$	-1	$	\\
				$m_{15}$	&	$	0	$	&	$	0	$	&	$	-1	$	&	$	1	$	&	$	-1	$	&	$	0	$	&	$	1	$	&	$	1	$	&	$	1	$	&	$	1	$	&	$	-1	$	&	$	0	$	\\

				\bottomrule
			\end{tabular}
		\end{table*}

\begin{figure}[hbt]
			\centering
			\psfrag{M1}{$\mathbf{M}_1$}
			\psfrag{M2}{$\mathbf{M}_2$}
			\includegraphics[width=10cm]{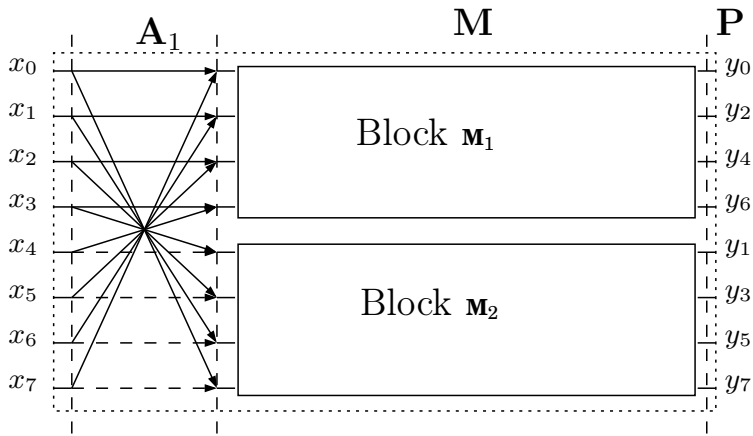}
			\caption{Signal flow graph for $\mathbf{T}_1$,  $\mathbf{T}_3$, and  $\mathbf{T}_{13}$.}
			\label{F:DiagramaT1}
		\end{figure}

\begin{figure}[hbt]
			\centering
			\psfrag{M1}{$\mathbf{M}_1$}
			\psfrag{M2}{$\mathbf{M}_2$}
			\includegraphics[width=10cm]{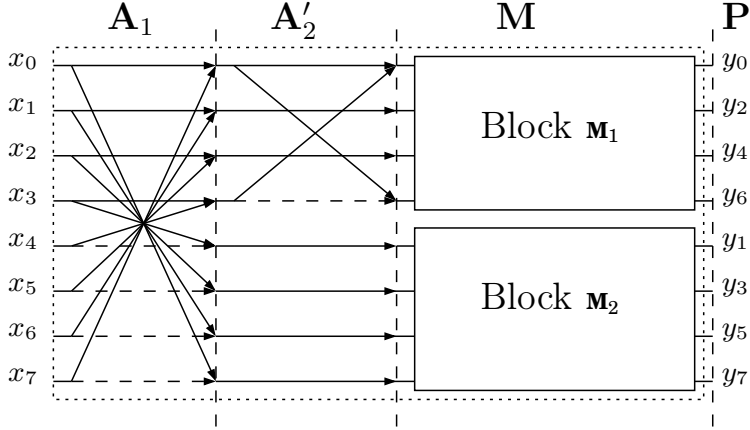}
			\caption{Signal flow graph for $\mathbf{T}_{16}$ and $\mathbf{T}_{17}$.}
			\label{F:DiagramaT16}
		\end{figure}

\begin{figure}[hbt]
			\centering
			\psfrag{M1}{$\mathbf{M}_1$}
			\psfrag{M2}{$\mathbf{M}_2$}
			\includegraphics[width=10cm]{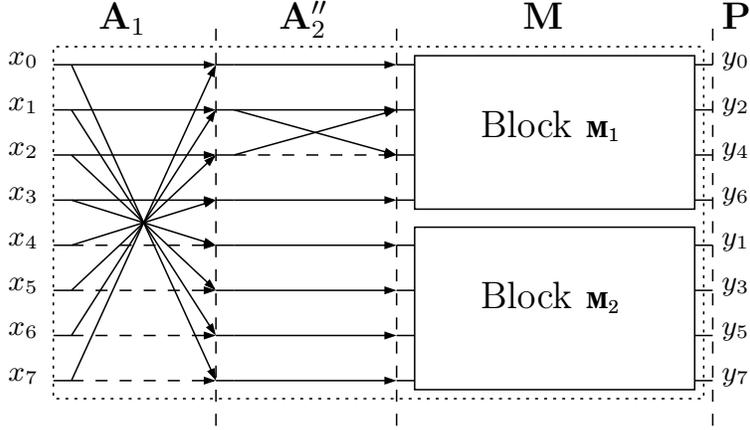}
			\caption{Signal flow graph for  $\mathbf{T}_{18}$.}
			\label{F:DiagramaT18}
		\end{figure}

\begin{figure}[hbt]
			\centering
			\psfrag{m0}[][][0.8]{$m_0$}
			\psfrag{m1}[][][0.8]{$m_1$}
			\psfrag{m2}[][][0.8]{$m_2$}
			\psfrag{m3}[][][0.8]{$m_3$}
			\psfrag{m4}[][][0.8]{$m_4$}
			\psfrag{m5}[][][0.8]{$m_5$}
			\psfrag{m6}[][][0.8]{$m_6$}
			\psfrag{m7}[][][0.8]{$m_7$}
			\psfrag{m8}[][][0.8]{$m_8$}
			\psfrag{m9}[][][0.8]{$m_9$}
			\psfrag{m10}[][][0.8]{$m_{10}$}
			\psfrag{m11}[][][0.8]{$m_{11}$}
			\psfrag{m12}[][][0.8]{$m_{12}$}
			\psfrag{m13}[][][0.8]{$m_{13}$}
			\psfrag{m14}[][][0.8]{$m_{14}$}
			\psfrag{m15}[][][0.8]{$m_{15}$}
			\psfrag{m16}[][][0.8]{$m_{16}$}
			\psfrag{m17}[][][0.8]{$m_{17}$}
			\includegraphics[width=10cm]{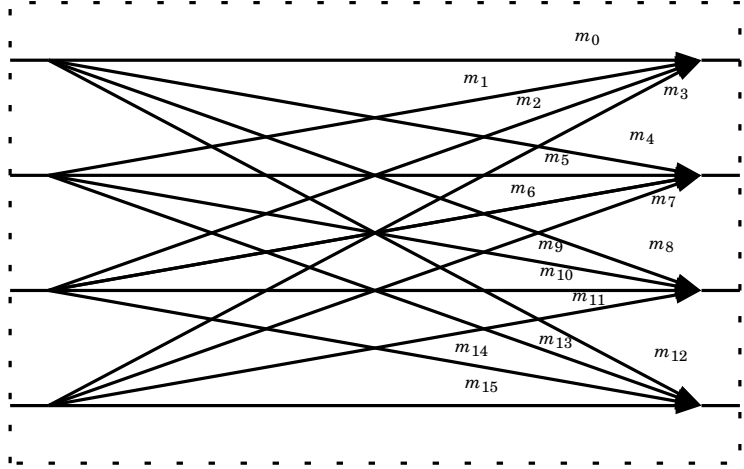}
			\caption{Blocks $\mathbf{M}_1$ and $\mathbf{M}_2$ from the signal flow graphs.}\label{f:blocos}
		\end{figure}

		Figs.~\ref{F:DiagramaT1}, \ref{F:DiagramaT16}, and \ref{F:DiagramaT18} present the signal flow graphs of the fast algorithms. Diagrams relate the input data $x_n$, $n = 0, 1, \ldots, 7$, to the output data $y_k$, $k = 0, 1, \ldots,7$,
		resulting in $\mathbf{y} = \mathbf{T}\cdot\mathbf{x}$. Here, dashed arrows represent multiplications by $-1$. When two or more arrows meet, their values are added~\cite{blahut2010fast}. Blocks $\mathbf{M}_1$ and $\mathbf{M}_2$ share the same structure in all diagrams except for the value of the constants presented in Table~\ref{t:emes} and are displayed in Fig.~\ref{f:blocos}.

		\subsection{Computational Complexity}
		The computational complexity of the proposed transforms can be estimated by the arithmetic complexity,
		given by the number of multiplications, addition and
		bit-shifting operations required for its implementation.
Table~\ref{t:complexidade} presents the arithmetic complexity of the discussed fast algorithms.
		In addition, we outline the computational cost involved in the direct computation of the $8$-point KLT, the $8$-point DCT utilizing the fast algorithms proposed by Loeffler~\cite{loeffler1989practical}, and the computational cost associated with the 8-point approximations for KLT as proposed in \cite{radunz2021data} and \cite{radunz2021low}.
		Compared to the SKLT and RKLT approximations, the proposed transforms exhibit an increase in the number of additions and bit-shifting. However, this rise is not a concern since our objective is to achieve multiplierless approximations.
		The quantities of additions and bit-shifting of the transforms $ \mathbf{T}_3$, $\mathbf{T}_{16}$, and $\mathbf{T}_{17}$ have additional factors because these transforms have $\pm 3$ elements in their matrices.
		From the proposed transforms, we can highlight $\mathbf{T}_1$, $\mathbf{T}_{13}$, and $\mathbf{T}_{18}$ as the ones that have the lower arithmetic cost,
		with a reduction of $57.14\%$, $53.57\%$, and $53.57\%$ in the number of additions in comparison with the exact KLT, respectively.
		These transforms perform well when compared with the fast algorithms for classical $8$-point low-complexity approximations such as the SDCT~\cite{haweel2001new},
		which requires only
		$24$
		addition operations for its implementation.

		\begin{table}[hbt]
			\caption{Comparison of the arithmetic complexity of the $8$-point transforms}
			\label{t:complexidade}
			\centering
			\begin{tabular}{lccc}
				\toprule
				Transform & Additions & Multiplications & Bit-shifts \\
				\midrule
				KLT & $56$ & $64$ & $0$ 	\\
				DCT~\cite{loeffler1989practical}  & $29$& $11$ & $0$ 	\\
				$\tilde{\mathbf{T}}_1$ (SKLT)~\cite{radunz2021data} & $24$& $0$ & $0$ \\
				$\tilde{\mathbf{T}}_2$ (SKLT)~\cite{radunz2021data} & $24$& $0$ & $0$ \\
				 $\widehat{\mathbf{T}}_1$ (RKLT)~\cite{radunz2021low} & $24$& $0$ & $0$  \\
				  $\widehat{\mathbf{T}}_2$ (RKLT)~\cite{radunz2021low} & $24$& $0$ & $0$  \\
				   $\widehat{\mathbf{T}}_3$ (RKLT)~\cite{radunz2021low} & $24$& $0$ & $0$  \\
				    $\widehat{\mathbf{T}}_4$ (RKLT)~\cite{radunz2021low,cintra2011dct} & $22$& $0$ & $0$  \\
				\addlinespace[1ex]
				\hdashline
				\addlinespace[1ex]
				$\mathbf{T}_1$ & $24$ & $0$ & $0$ 	\\
				$\mathbf{T}_3$ & $30 + 18$ & $0$ & $6 + 18$ 	\\
				$\mathbf{T}_{13}$& $26$ & $0$ & $13$ 	\\
				$\mathbf{T}_{16}$& $28 + 10$ & $0$ & $12 + 10$  	\\
				$\mathbf{T}_{17}$& $27 + 11$ & $0$ & $11 + 11$  	\\
				$\mathbf{T}_{18}$& $26$ & $0$ & $12$ 	\\
				\addlinespace[1ex]
				\bottomrule
			\end{tabular}
		\end{table}

		\section{Experiments on Image Compression}\label{s:imagecomp}
		\subsection{JPEG-like Compression}
		The compression performance of the proposed transforms can be evaluated when applied to image coding experiments, as well as in~\cite{cintra2011dct, bouguezel2008low, cintra2014low}.
		For simplicity, but without loss of generality, $8$-bit images in gray scale were considered.
		The JPEG-like compression methodology used in this experiment is presented as follows~\cite{salomon2004data}. The input image was divided into disjoint $8 \times 8$ sub-blocks. Let $\mathbf{A}$ be a sub-block. The direct two-dimensional (2D) transform was applied in each sub-block, resulting in $\mathbf{B} = \widehat{\mathbf{K}} \cdot \mathbf{A} \cdot \widehat{\mathbf{K}}^{\top}$~\cite{suzuki2010integer}.
		Considering the zig-zag pattern~\cite{salomon2004data}, the initial $r$ coefficients from $\mathbf{B}$ were retained, resulting in truncated sub-blocks $\mathbf{\bar{B}}$. The 2D inverse transform was applied in each sub-block $\mathbf{\bar{B}}$, resulting in $\mathbf{\bar{A}} = \widehat{\mathbf{K}}^{-1} \cdot \mathbf{\bar{B}} \cdot (\widehat{\mathbf{K}}^{-1})^\top$. The compressed sub-blocks $\mathbf{\bar{A}}$ were recomposed in the place of the originals sub-blocks $\mathbf{A}$. Finally, the compressed image was compared to the original image to evaluate the loss of quality imposed by compression.

		For assessing the quality of compressed images, we used as figures of merit
		the peak signal-to-noise ratio (PSNR)~\cite{Huynh-Thu2008Scope} and the mean structural similarity index (MSSIM)~\cite{wang2004image}.
		Even though it is a very popular metric, it was shown in~\cite{wang2009mean} that the PSNR is not the best measure when it comes to predict the human perception of image quality~\cite{wang2004image,wang2009mean}. Nevertheless, we considered this figure of merit for comparison purposes.
		On the other hand, the MSSIM was shown to be capable of closely
		capturing the image quality as understood by the human visual system model~\cite{wang2004image}.

		\subsection{Results and Discussion}
		In this subsection, we present the outcomes achieved by employing the proposed transforms in the context of image compression. To facilitate comparison, we included assessments of the KLT approximations developed in \cite{radunz2021data} and \cite{radunz2021low}, along with evaluations of the exact DCT and KLT. The findings underscore the significance of the approximations introduced in this study, as one of the approximations outperformed both the DCT and KLT for a specific image.

\begin{figure}[t]
			\centering
			\subfloat[\textit{Lena}]{\label{f:lenaoriginal}
				\includegraphics[width=5cm]{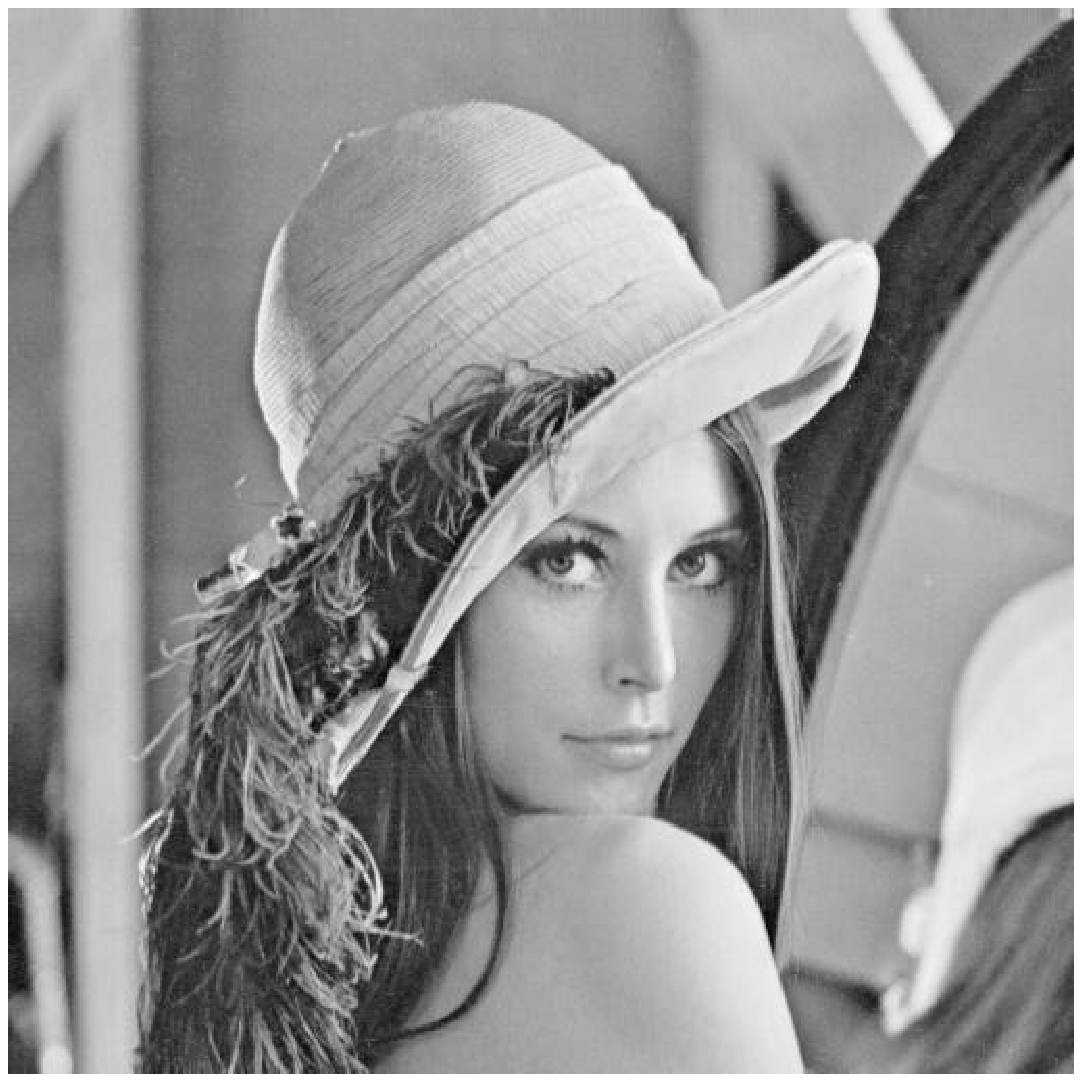}}
			\subfloat[\textit{Grass}]{\label{f:grassorig}
				\includegraphics[width=5cm]{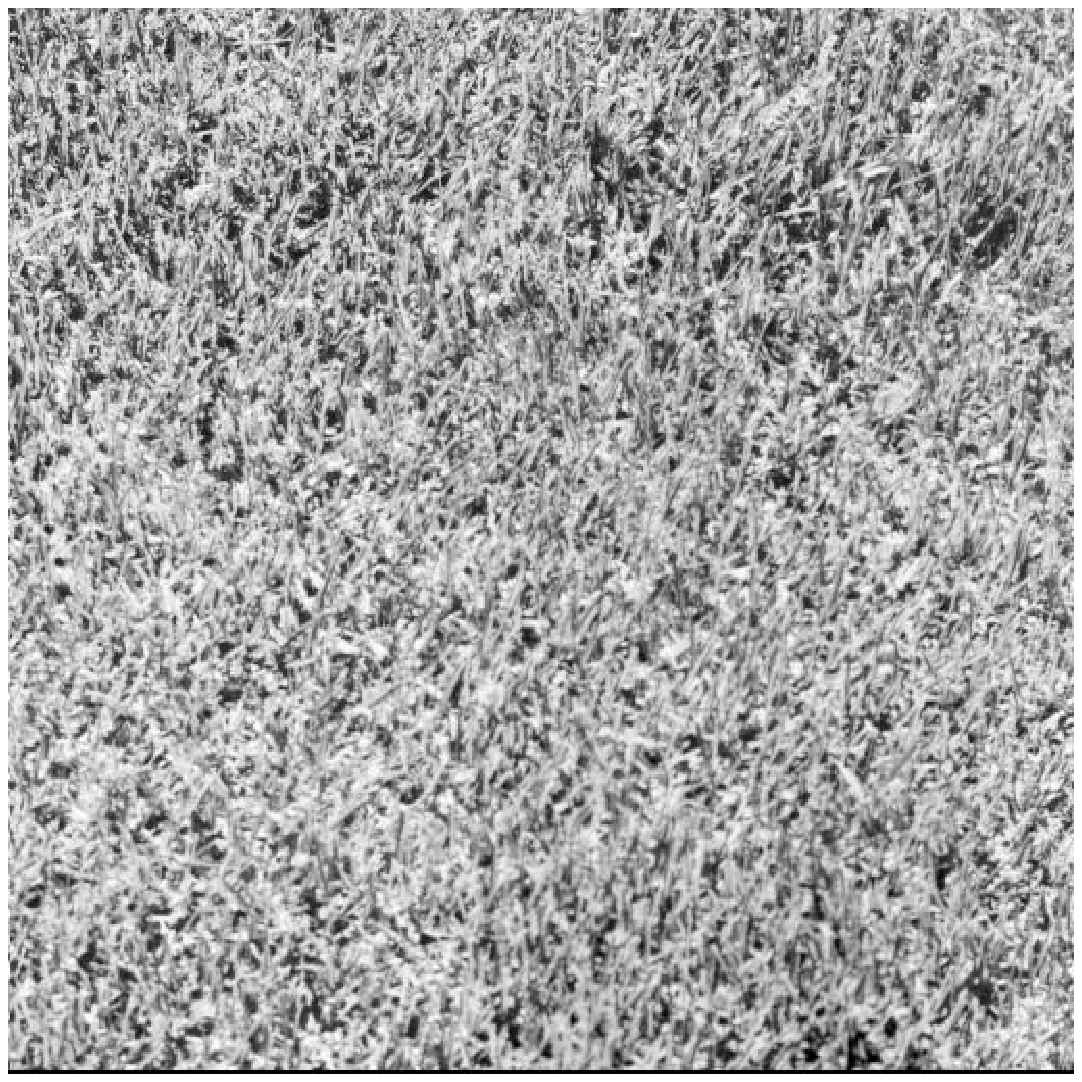}}
			\caption{Original Images.}\label{f:ImagensOriginais}
		\end{figure}

\begin{figure}[t]
			\centering
			\subfloat[\textit{$\widehat{\mathbf{K}}_1$}]{\label{f:lenaT1}
				\includegraphics[width=0.25\linewidth]{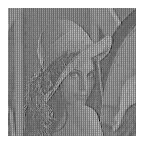}}
			\subfloat[$\widehat{\mathbf{K}}_3$]{\label{f:lenaT2}
				\includegraphics[width=0.25\linewidth]{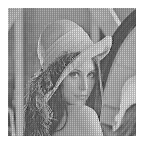}}
			\subfloat[\textit{$\widehat{\mathbf{K}}_{13}$ }]{\label{f:lenaT3}
				\includegraphics[width=0.25\linewidth]{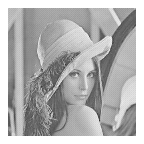}}
			\subfloat[\textit{$\mathbf{K}^{(0.2)}$ }]{\label{f:lenaKLT02}
				\includegraphics[width=0.25\linewidth]{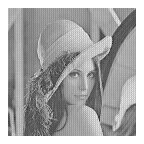}}

			\subfloat[$\widehat{\mathbf{K}}_{16}$]{\label{f:lenaT4}
				\includegraphics[width=0.25\linewidth]{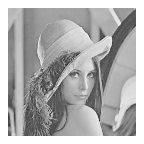}}
			\subfloat[\textit{$\widehat{\mathbf{K}}_{17}$}]{\label{f:lenaT5}
				\includegraphics[width=0.25\linewidth]{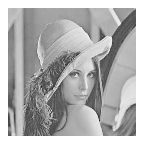}}
			\subfloat[$\widehat{\mathbf{K}}_{18}$]{\label{f:lenaT6}
				\includegraphics[width=0.25\linewidth]{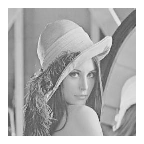}}
			\subfloat[\textit{$\mathbf{K}^{(0.8)}$ }]{\label{f:lenaKLT08}
				\includegraphics[width=0.25\linewidth]{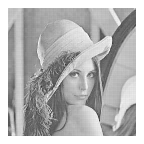}}
			\caption{Compressed \textit{Lena} Images.}\label{f:compressedLena}
		\end{figure}

\begin{figure}[t]
			\centering
			\subfloat[\textit{$\widehat{\mathbf{K}}_1$ }]{\label{f:grassT1}
				\includegraphics[width=0.25\linewidth]{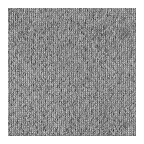}}
			\subfloat[$\widehat{\mathbf{K}}_3$ ]{\label{f:grassT2}
				\includegraphics[width=0.25\linewidth]{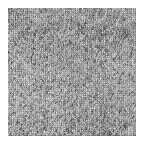}}
			\subfloat[\textit{$\widehat{\mathbf{K}}_{13}$ }]{\label{f:grassT3}
				\includegraphics[width=0.25\linewidth]{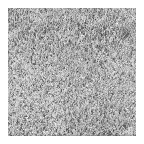}}
			\subfloat[\textit{$\mathbf{K}^{(0.2)}$ }]{\label{f:grassKLT02}
				\includegraphics[width=0.25\linewidth]{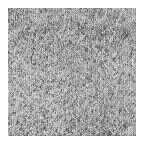}}

			\subfloat[$\widehat{\mathbf{K}}_{16}$ ]{\label{f:grassT4}
				\includegraphics[width=0.25\linewidth]{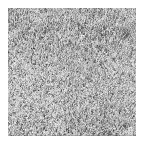}}
			\subfloat[\textit{$\widehat{\mathbf{K}}_{17}$ }]{\label{f:grassT5}
				\includegraphics[width=0.25\linewidth]{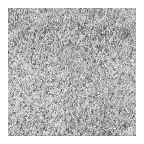}}
			\subfloat[$\widehat{\mathbf{K}}_{18}$ ]{\label{f:grassT6}
				\includegraphics[width=0.25\linewidth]{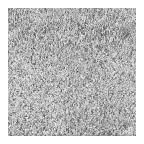}}
			\subfloat[\textit{$\mathbf{K}^{(0.8)}$ }]{\label{f:grassKLT08}
				\includegraphics[width=0.25\linewidth]{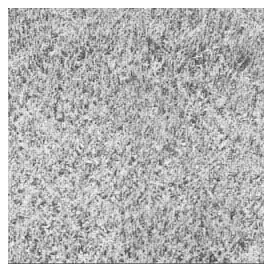}}
			\caption{Compressed \textit{Grass} Images.}\label{f:compressedGrass}
		\end{figure}

		Fig.~\ref{f:ImagensOriginais} presents the original \textit{Lena} and \textit{Grass}
		images~\cite{sipi2005usc} used in the qualitative analysis.
		In this step, each image was submitted to a compression rate (CR) of $85\%$,
		$r = 10$. Figs.~\ref{f:compressedLena} and \ref{f:compressedGrass} present the compressed images using the proposed transforms and the exact KLT for $\rho = 0.2$ and $0.8$.

		Table~\ref{t:PSNRMSSIM} presents the  PSNR and MSSIM values for the compressed images.
		 In addition to the values considering the proposed transforms, we included the values for the KLT approximations proposed in \cite{radunz2021data} and \cite{radunz2021low}, the exact KLT for $\rho = 0.95$, and the exact DCT. %
		The proposed transforms perform well, and in some cases even better than the exact KLT, for a given value of $\rho$ within the interval of each group of transforms.
		We highlighted the values of the best measurements for each group of the approximate transforms.
		Particular emphasis is placed on the superiority of the approximations $\widehat{\mathbf{K}}_{13}$, $\widehat{\mathbf{K}}_{16}$, and $\widehat{\mathbf{K}}_{18}$ which have demonstrated superior performance compared to the known KLT approximations documented in the literature.
		The proposed transforms $\widehat{\mathbf{K}}_1$, $\widehat{\mathbf{K}}_3$, and  $\widehat{\mathbf{K}}_{13}$ were derived considering low values of $\rho$. As the pixels of a natural image are highly correlated~\cite{rao2000transform}, image compression using these transforms does not show the best results, as expected.
		However,
		{we
			can emphasize
			that $\widehat{\mathbf{K}}_{16}$
			outperformed
			the exact KLT and DCT
			considering the \textit{Grass} image.}
		The values
		are highlighted
		in the table with a red box.
		Also,
		considering
		the other proposed transforms
		we
		can see
		that, qualitatively, there is no visually perceptible differences between the compressed images considering the approximate transforms and the exact KLT.

		\begin{table}[t] \caption{Image quality measures}
			\label{t:PSNRMSSIM}
			\centering
				\begin{tabular}{llllllllllll}
					\toprule
					Image                          & \multicolumn{2}{c}{\textit{Lena}}                                                                                                      & \multicolumn{2}{c}{\textit{Grass}}
					\\
					\midrule
					Transform
					& \multicolumn{1}{c}{PSNR} & \multicolumn{1}{c}{MSSIM} &
					\multicolumn{1}{c}{PSNR} & \multicolumn{1}{c}{MSSIM} \\
					\midrule

					$\widehat{\mathbf{K}}_1$
					& $10.7230$                & $0.1086$
					& $10.2617$                & $0.3440$    \\
					$\widehat{\mathbf{K}}_3$
					& $18.2075$&$0.2698$
					& $16.1770$     & $0.6216$      \\
					$\widehat{\mathbf{K}}_{13}$
					& $\textbf{30.5265}$& $\textbf{0.8093}$
					& $\textbf{19.6360}$  & $\textbf{0.7797}$   \\
					$\mathbf{K}^{(0.2)}$
					& $20.14173$ & $0.3302$
					& $17.3274$ & $0.6759$ \\
                    $\tilde{\mathbf{T}}_1$ (SKLT)
					& $26.5803$  & $0.8293$
					& $16.5951$ & $ 0.6585$  \\
					$\widehat{\mathbf{T}}_1$ (RKLT)
					& $10.7230$  & $0.1086$
					& $10.2617$ & $0.3440$   \\
						$\widehat{\mathbf{T}}_2$ (RKLT)
					& $23.6696$  & $0.4605$
					& $17.8777$ & $0.6997$   \\
					\addlinespace[1ex]
					\hdashline
					\addlinespace[1ex]
					$\widehat{\mathbf{K}}_{16}$
					& $\textbf{31.8353}$ & $0.8942$
					& $\textbf{19.9568}$  & $\textbf{0.7884}$  \\
					$\widehat{\mathbf{K}}_{17}$
					& $31.7447$ & $0.8934$
					& $19.9213$     & $0.7874$     \\
					$\widehat{\mathbf{K}}_{18}$
					& $31.6908$ & $\textbf{0.9091}$
					& $19.59472$ & $0.7776$  \\
					$\mathbf{K}^{(0.8)}$
					& $29.9278$ & $0.7584$
					& $19.8954$ & $0.7861$ \\
					$\tilde{\mathbf{T}}_2$ (SKLT)
					& $27.4416$  & $0.8577$
					& $17.0181$ & $ 0.6777$  \\
					$\widehat{\mathbf{T}}_3$ (RKLT)
					& $22.9120$  & $0.4236$
					& $17.6256$ & $0.6869$  \\
					$\widehat{\mathbf{T}}_4$ (RKLT)
					& $30.4424$  & $0.8932$
					& $19.1573$ & $ 0.7425$  \\
					\addlinespace[1ex]
					\hdashline
					\addlinespace[1ex]
					$\mathbf{K}^{(0.95)}$
					& $31.9935$   & $0.9019$
					& $19.9384$    & $0.7864$    \\
					DCT
					& $32.0814$  & $0.9136$
					& $19.893$ & $0.7839$   \\

					\bottomrule
				\end{tabular}
		\end{table}

\begin{figure}[t]
			\centering
			\subfloat[Average PSNR considering $C_1$ group approximate transforms]{
				\includegraphics[width=0.45\linewidth]{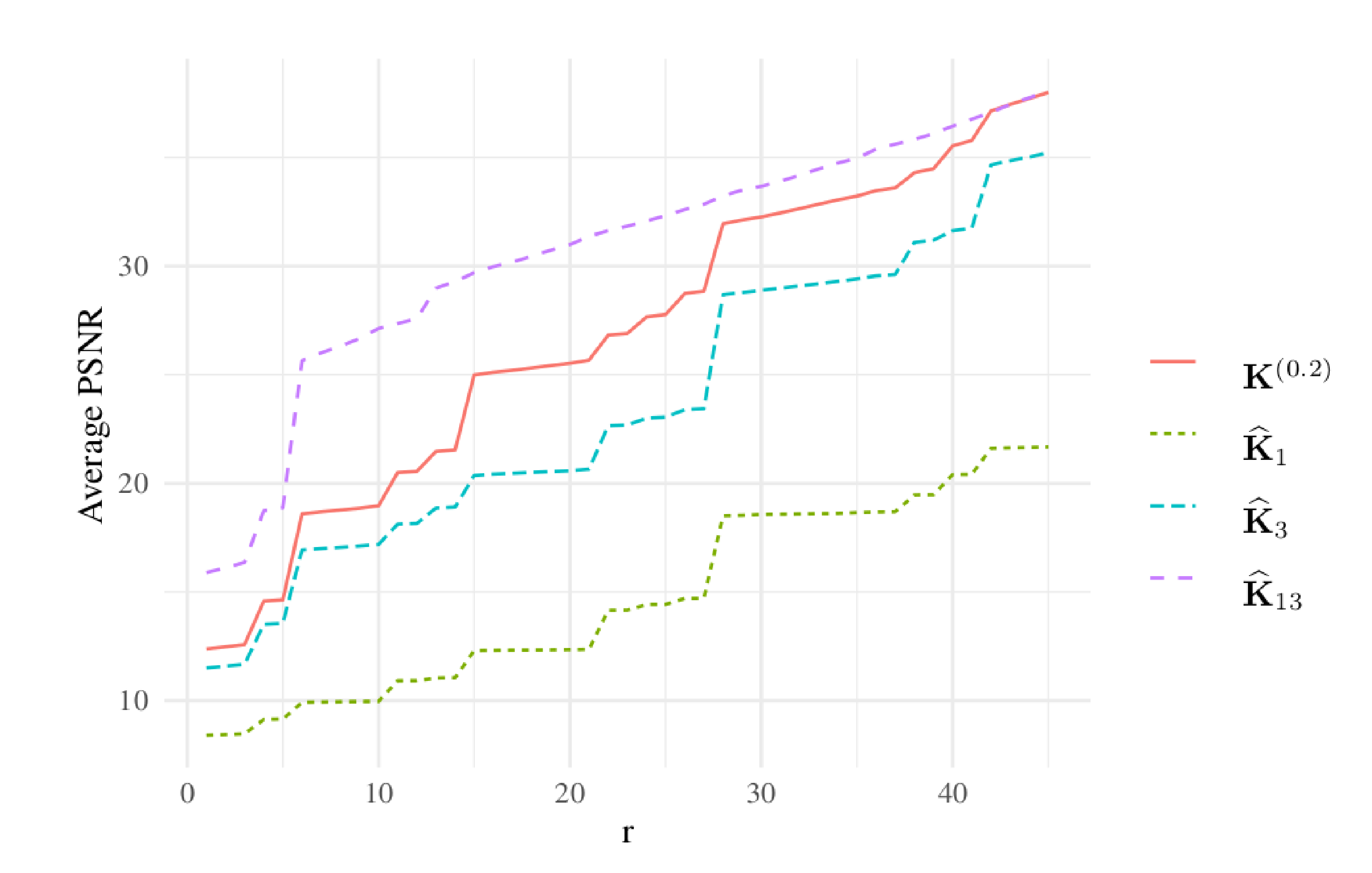}}
			\subfloat[Average PSNR considering $C_2$ group approximate transforms]{
				\includegraphics[width=0.45\linewidth]{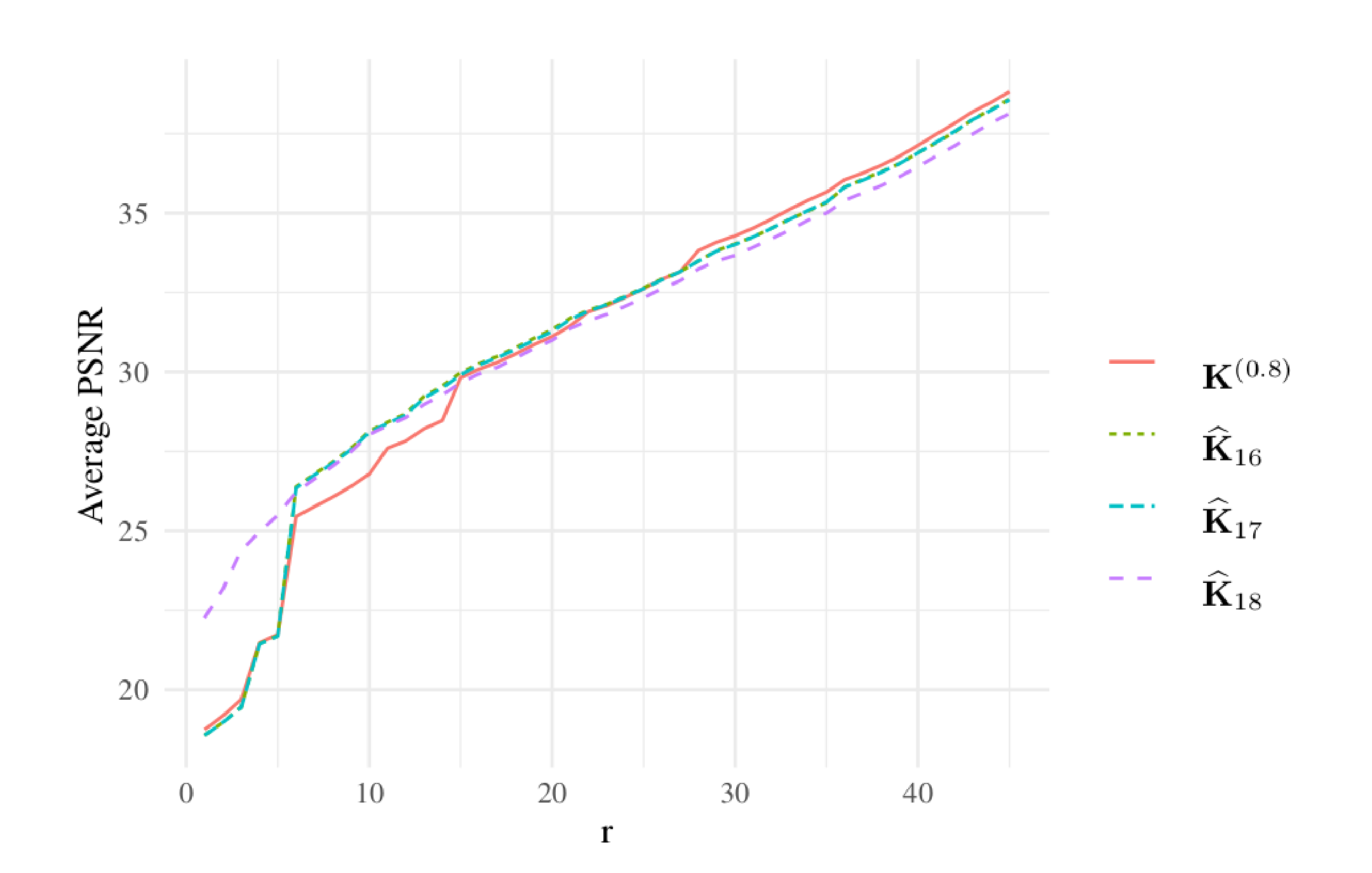}} \\
			\subfloat[Average MSSIM considering $C_1$ group approximate transforms]{
				\includegraphics[width=0.45\linewidth]{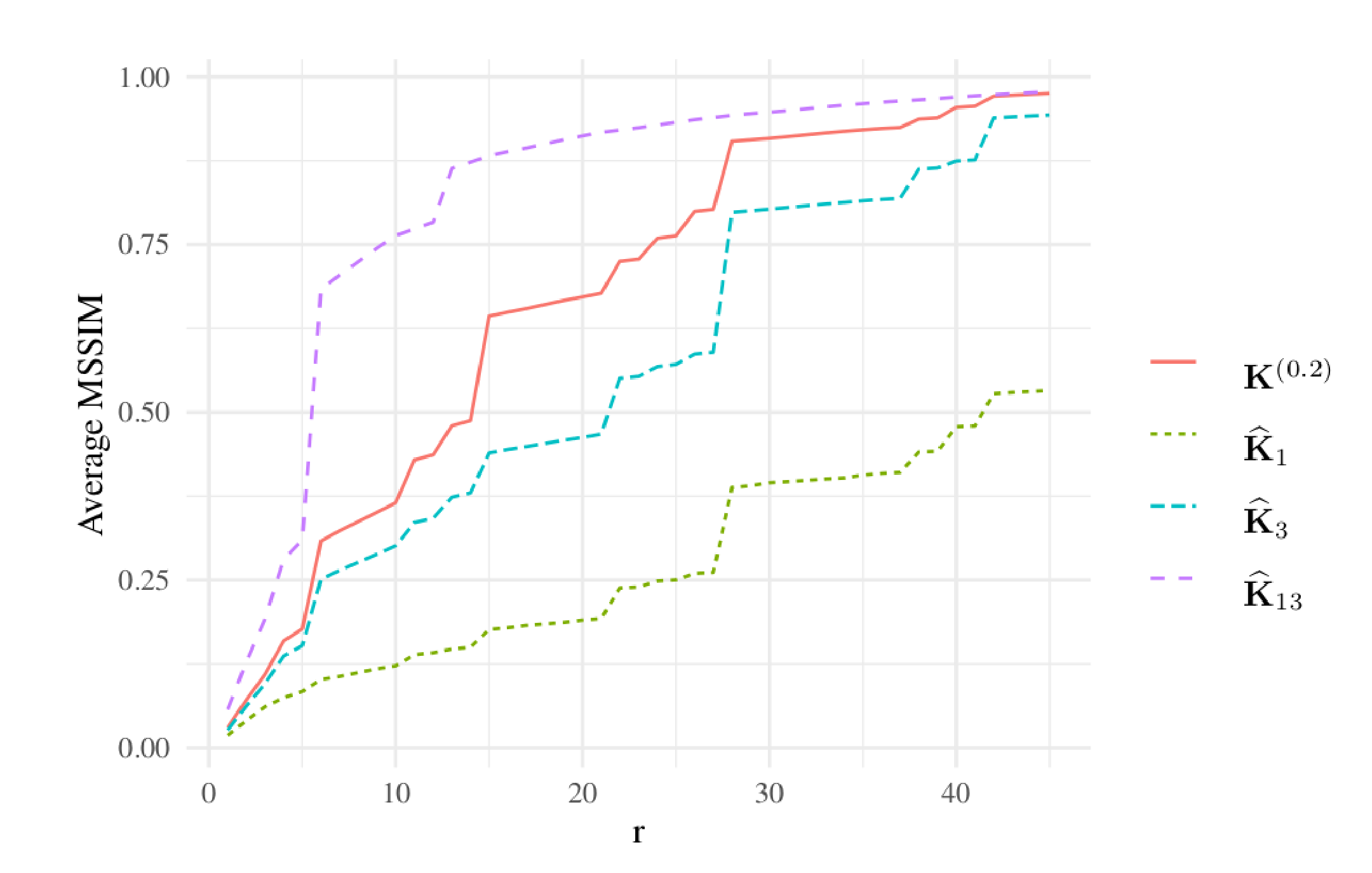}}
			\subfloat[Average MSSIM considering $C_2$ group approximate transforms]{
				\includegraphics[width=0.45\linewidth]{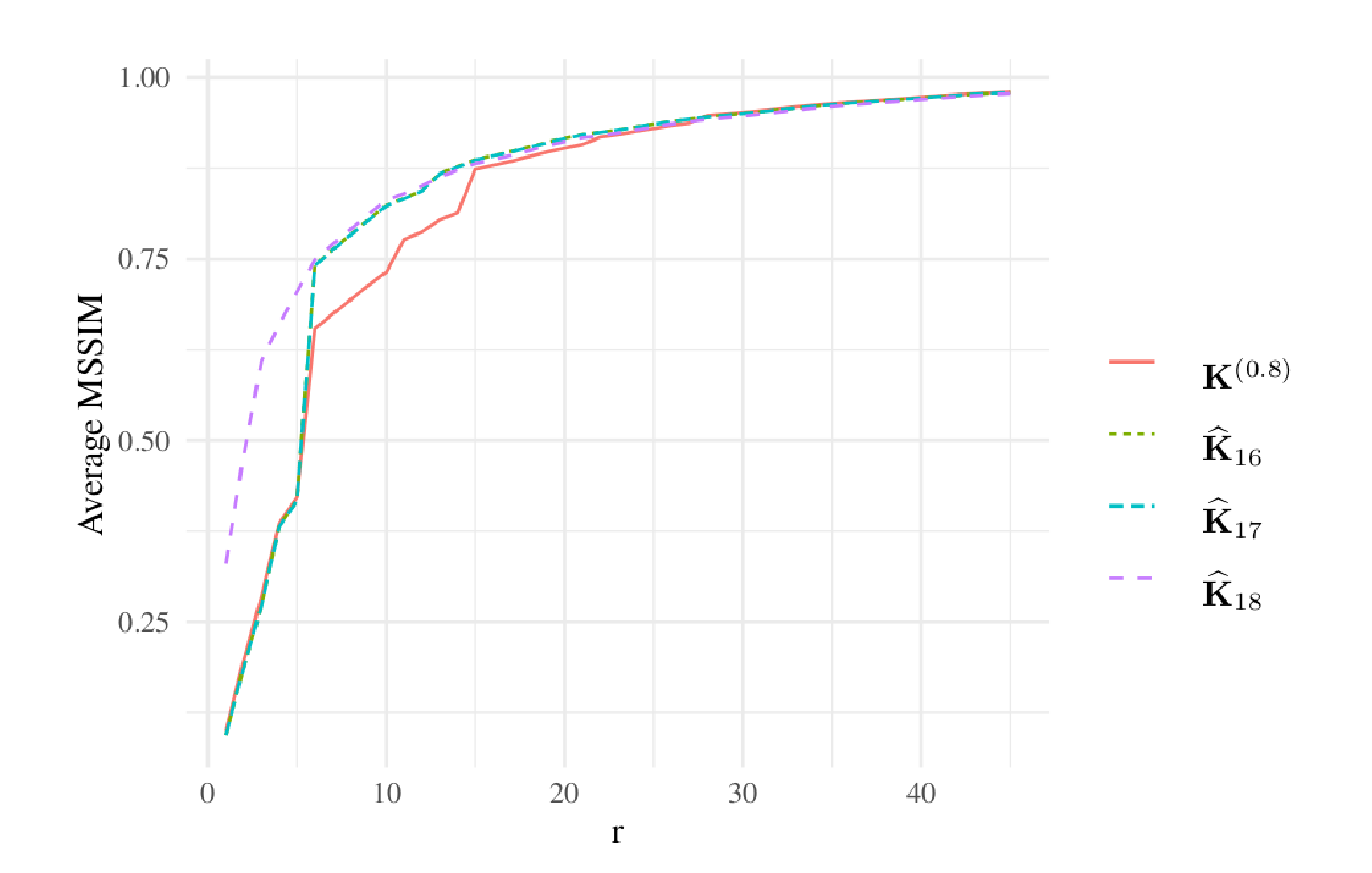}}
			\caption{Quality measures of the considered approximations for several	values of $r$ according to the figures of merit.}\label{f:medidasimagensPSNR}
		\end{figure}

		We extended the experiment to a group of $45$ $512 \times 512$ $8$-bit greyscale images, obtained from~\cite{sipi2005usc}, considering different rates of compression ($1 \leq r \leq 45$). The PSNR and MSSIM measures were computed for each image, and the average of these values were taken. Fig.~\ref{f:medidasimagensPSNR} presents the plots of the average values of these measures.
		There are two graphs for each figure of merit, one for each group of the approximate transforms, $C_1$ and $C_2$.
		In order to compare the approximate transforms we also calculated this measurements for the exact KLT considering the values of $\rho = 0.2$ and $0.8$. The proposed transforms performed very well when compared with the exact KLT, and considering the transforms from group $C_2$ {they outperformed the exact KLT} for $0<r<15$ approximately.

		\section{Hardware Implementation} \label{sec:hardware}

		The 8-point low-complexity
		transforms outlined in Table~\ref{t:transforms}
		were implemented on an
		FPGA.
		The platform adopted for the hardware
		implementation was
		the Xilinx Artix-7
		XC7A35T-1CPG236C.
		Notice we do not implement the diagonal elements of the approximations.
		This is because they can be easily incorporated in the quantization step in image and video compression schemes~\cite{wallace1992jpeg}.

		The designs were implemented using a
		pipelined systolic architecture
		for each of the transforms~\cite{Baghaie2000, Safiri1996} considering 8-bit wordlength inputs.
		Each transform implementation
		is split in different sub-blocks.
		Each sub-block implements a
		different matrix in its corresponding fast algorithm as in~\eqref{eq:f1},~\eqref{eq:f2}, and~\eqref{eq:f3}
		and displayed in Fig.~\ref{F:DiagramaT1}, Fig.~\ref{F:DiagramaT16}, and Fig.~\ref{F:DiagramaT18}, respectively.
		Each sub-block that requires an arithmetic operation expands
		the wordlength in one bit in order to avoid overflow.
		The sub-block implementing the permutation matrix~$\mathbf{P}$ in~\eqref{eq:P}
		contains only combinational logic as it only requires
		re-routing of the transform coefficients and does not possess any arithmetic operation.
		The kernel~$\mathbf{M}$ of all the transforms in~\eqref{eq:Ki}
		are implemented with two clock cycles of latency.
		This is because each row of each of the transform kernel
		possesses at least three nonzero entries~(cf.~Table~\ref{t:emes}).
		The intermediary matrices~$\mathbf{A}_1$,~$\mathbf{A}_2^{\prime}$, and~$\mathbf{A}_2^{\prime\prime}$
		in the factorizations in~\eqref{eq:f1},~\eqref{eq:f2}, and~\eqref{eq:f2}, respectively,
		require at most an addition of two elements per row, therefore being enough only one clock cycle to implement each.

\begin{figure}[t]
			\centering
			\includegraphics[scale=0.7]{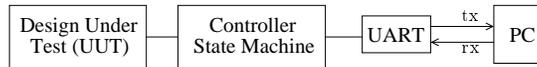}
			\caption{Testbed architecture for testing the implemented designs.}
			\label{fig:testbed}
		\end{figure}

		The designs were implemented and tested according to
		the scheme shown in Fig.~\ref{fig:testbed},
		along with a state-machine serving as controller and connected to a
		universal asynchronous receiver-transmitter (UART) block.
		The UART core interfaces with the controller
		through an ARM Advanced Microcontroller Bus Architecture Advanced eXtensible Interface 4
		(AMBA AXI-4) protocol.

		The personal computer (PC)
		communicates with the controller through the UART by
		sending a set of eight 8-bit coefficients,
		which corresponds to an input for the transform block under test.
		The values of the 8-bit coefficients are drawn from a uniform distribution in the interval~$[-10, 10]$.
		The set of the eight coefficients are then sent to the design
		and processed.
		After processed,
		the controller
		sends the eight output coefficients
		back to the PC, which is compared with the
		output of a software model used to ensuring
		the hardware design is accurately implemented.

		Table~\ref{t:hardware}
		shows the hardware resources utilization and metrics for the transforms
		in Table~\ref{t:transforms}.
		The considered figures of merit are the
		number of occupied slices,
		number of look-up tables~(LUT),
		flip-flop~(FF) count,
		wordlength increase ($\Delta~\text{\#bits}$),
		latency ($L$) in terms of clock cycles,
		critical path delay~($T_{\text{cpd}}$),
		maximum operating frequency $F_{\text{max}} = T_{\text{cpd}}^{-1}$,
		and dynamic power~($D_p$) normalized by~$F_{\text{max}}$.

		\begin{table}[t]
			\centering
			\caption{FPGA measures of the implemented architectures for the new and competing transforms}
			\begin{tabular}{l@{\enskip }c@{\enskip }c@{\enskip }c@{\enskip }c@{\enskip }c@{\enskip }c@{\enskip }c@{\enskip }c}
				\toprule
				\multirow{3}{*}{Transform} & \multicolumn{8}{c}{Metrics} \\ \cline{2-9}
				& \multirow{2}{*}{Slices} & \multirow{2}{*}{LUT} & \multirow{2}{*}{FF} & $\Delta$ & $L$ & $T_{\text{cpd}}$ & $F_{\text{max}}$& $D_p$ \\
				& & & & \text{\#bits} & (cycles) & ($\nano\second$) & ($\mega\hertz$) & ($\mu\watt\per\mega\hertz$) \\\midrule
				$\mathbf{T}_{1}$ & \textbf{75} & \textbf{217} & \textbf{279} & \textbf{3} & \textbf{3} &  \textbf{3.691} & \textbf{270.929} & \textbf{33.219}\\
				$\mathbf{T}_{3}$ & 150 & 471 & 370 & 5 & \textbf{3} &  4.961 & 201.572 & 54.571\\
				$\mathbf{T}_{13}$ & 93 & 277 & 334 & 4 & \textbf{3} &  4.203 & 237.925 & 37.827\\
				$\mathbf{T}_{16}$ & 143 & 406 & 444 & 6 & 4 &  4.926 & 203.004 & 54.186\\
				$\mathbf{T}_{17}$ & 148 & 426 & 444 & 6 & 4 &  5.072 & 197.161 & 55.792\\
				$\mathbf{T}_{18}$ & 110 & 287 & 401 & 5 & 4 &  4.580 & 218.341 & 41.220\\\bottomrule
			\end{tabular}{\label{t:hardware}}
		\end{table}

		Among the considered transforms, the $\mathbf{T}_{1}$ is the
		one requiring the least amount of resources such as FFs, LUT, and, consequently, slices.
		This is due to two factors: (i) the latency~$L$, and (ii) wordlength increase~$\Delta~\text{\#bits}$.
		Smaller latency means less registers are needed for storing information, directly reducing the need for FFs and LUTs.
		Also, with reduced wordlength increase~$\Delta~\text{\#bits}$,
		less routing resources are needed inside the device to attain the desired computation.

		The latency is a direct consequence of the fast algorithm
		that is found for the considered transforms, as outlined in~\eqref{eq:f1},~\eqref{eq:f2}, and~\eqref{eq:f3}.
		The transform~$\mathbf{T}_{1}$ is factorized with the simplest of the fast algorithms,
		involving only three matrices --- in fact only two that requires arithmetic operations --- as compared to the other algorithms that requires one more matrix --- in fact only three that requires arithmetic operations.
		An inspection of Table~\ref{t:emes} along with~\eqref{eq:Ki}
		also shows that the transform kernel~$\mathbf{M}$ for~$\mathbf{T}_{1}$
		requires only additions of at most three elements per row.
		It does not require any constant multiplication by two (a bit-shifting operation) or by three (a bit-shifting plus addition)
		like the other transforms in Table~\ref{t:transforms}~(cf.~Table~\ref{t:emes}), which then renders a transform requiring a smaller bit increment when compared to other proposed transforms.
		Because of the reduced amount of resources
		when compared to the other transforms in Table~\ref{t:hardware},~$\mathbf{T}_{1}$ is
		also the transform with the least critical path delay,
		and therefore the highest maximum operating frequency and normalized dynamic power.

		The transform~$\mathbf{T}_{13}$
		is the second most economical in terms of hardware resources considering FFs, LUTs, and slices.
		It shares in common with~$\mathbf{T}_{1}$ its fast algorithm,
		however, $\mathbf{T}_{13}$ kernel requires multiplications by
		constants as shown in~Table~\ref{t:emes},
		demanding a higher wordlength increment and therefore more resources than~$\mathbf{T}_{1}$.
		The transform~$\mathbf{T}_{17}$
		is the one requiring the most amount of resources and possessing the highest critical path delay,
		resulting in the lowest maximum operating frequency and normalized dynamic power in comparison to the other transforms in
		Table~\ref{t:hardware}.

		\section{Conclusions}\label{S:conclusion}

		In this paper,
		we proposed a new class of data-independent low-complexity KLT approximations.
		In prior studies, KLT approximations were devised using specific rounding functions, such as the Signed KLT (SKLT) and Rounded KLT (RKLT).
		To the best of our knowledge, the existing literature lacked low-complexity approximation transforms covering the entire correlation scenario ($0<\rho<1$). Leveraging this novelty and acknowledging the previously proposed KLT approximations' favorable balance between cost and performance, we embarked on an extended exploration for approximations, encompassing a broader range of rounding functions. This expanded search yielded additional candidates for approximations. After refining the results, we successfully identified optimal KLT approximations for different intervals of the correlation.

		The obtained approximations were derived applying a set of rounding functions to the elements of the exact KLT, varying the value of the correlation coefficient $\rho$. An optimization problem was solved aiming at the proposition of optimal transforms according to defined figures of merit. The $k$-means clustering method was used to classify the optimal transforms into groups to certain $\rho$ values intervals. Fast algorithms were derived for the optimal approximations proposed by factorizing the transforms matrices into sparse matrices. Only addition and bit-shifting operations were necessary for the implementation of the proposed transforms.
		Through the transform factorization, we managed to achieve a remarkable reduction of approximately $58\%$ in the arithmetic cost of $\mathbf{T}_1$ compared to the exact KLT.
		The applicability
		of the proposed approximation in the context of image compression was demonstrated. Our experiments showed that the proposed transforms performed very well when compared to the exact KLT, and in the cases of $\widehat{\mathbf{K}}_{16}$, $\widehat{\mathbf{K}}_{17}$, and $\widehat{\mathbf{K}}_{18}$ even \textbf{outperformed the exact KLT and DCT.}
		Acknowledging that our coverage spans the entire correlation scenario and recognizing the high correlation inherent in natural images, we anticipated limitations in identifying suitable applications for the transforms proposed for low values of $\rho$ ($\rho < 0.7$). In future works, our objective is to explore signals characterized by low correlation, thereby demonstrating alternative applications for these transforms.
		In addition to its application in image compression, we explored FPGA hardware implementation, showing a trade-off between performance and resource usage and performance,
		where the~$\mathbf{T}_{1}$ requires the least amount of resources and~$\mathbf{T}_{17}$ the highest amounts of FFs and LUTs.

\onecolumn

{\small
\singlespacing
\bibliographystyle{siam}
\bibliography{ref}

\begin{thebibliography}{10}

\bibitem{Coutinho2017}
{\sc V.~A.~Coutinho, R.~J. Cintra, and F.~M. Bayer}, {\em Low-complexity
  multidimensional {DCT} approximations for high-order tensor data
  decorrelation}, IEEE Transactions on Image Processing, 26 (2017),
  pp.~2296--2310.

\bibitem{ahmed1974discrete}
{\sc N.~Ahmed, T.~Natarajan, and K.~R. Rao}, {\em Discrete cosine transfom},
  IEEE Transactions on Computers, C-23 (1974), p.~90–93.

\bibitem{Baghaie2000}
{\sc R.~Baghaie and V.~Dimitrov}, {\em Computing {Haar} transform using
  algebraic integers}, Conference Record of Thirty-Fourth Asilomar Conference
  on Signal, Systems and Computers, 1 (2000), pp.~438--442.

\bibitem{bayer2012dct}
{\sc F.~M. Bayer and R.~J. Cintra}, {\em {DCT}-like transform for image
  compression requires 14 additions only}, Electronics Letters, 48 (2012),
  p.~919–921.

\bibitem{bhairannawar2020fpga}
{\sc S.~S. Bhairannawar, S.~Sarkar, and K.~Raja}, {\em {FPGA} implementation of
  optimized {K}arhunen--{L}o{\`e}ve transform for image processing
  applications}, Journal of Real-Time Image Processing, 17 (2020),
  pp.~357--370.

\bibitem{biswas2010improved}
{\sc M.~Biswas, M.~R. Pickering, and M.~R. Frater}, {\em Improved
  \mbox{{H}.264}-based video coding using an adaptive transform}, in 2010 IEEE
  International Conference on Image Processing, IEEE, 2010, p.~165–168.

\bibitem{blahut2010fast}
{\sc R.~E. Blahut}, {\em Fast algorithms for signal processing}, Cambridge
  University Press, Cambrigde, UK, 2010.

\bibitem{blanes2012divide}
{\sc I.~Blanes, J.~Serra-Sagrista, M.~W. Marcellin, and J.~Bartrina-Rapesta},
  {\em Divide-and-conquer strategies for hyperspectral image processing: A
  review of their benefits and advantages}, IEEE Signal Processing Magazine, 29
  (2012), pp.~71--81.

\bibitem{bouguezel2008low}
{\sc S.~Bouguezel, M.~O. Ahmad, and M.~Swamy}, {\em Low-complexity 8$\times$8
  transform for image compression}, Electronics Letters, 44 (2008),
  p.~1249–1250.

\bibitem{brahimi2020novel}
{\sc N.~Brahimi, T.~Bouden, T.~Brahimi, and L.~Boubchir}, {\em A novel and
  efficient 8-point {DCT} approximation for image compression}, Multimedia
  Tools and Applications, 79 (2020), pp.~7615--7631.

\bibitem{britanak2010discrete}
{\sc V.~Britanak, P.~C. Yip, and K.~R. Rao}, {\em Discrete cosine and sine
  transforms: general properties, fast algorithms and integer approximations},
  Academic Press, San Diego, CA, 2010.

\bibitem{cagnazzo2006low}
{\sc M.~Cagnazzo, L.~Cicala, G.~Poggi, and L.~Verdoliva}, {\em Low-complexity
  compression of multispectral images based on classified transform coding},
  Signal Processing: Image Communication, 21 (2006), pp.~850--861.

\bibitem{canterle2020multiparametric}
{\sc D.~R. Canterle, T.~L. da~Silveira, F.~M. Bayer, and R.~J. Cintra}, {\em A
  multiparametric class of low-complexity transforms for image and video
  coding}, Signal Processing, 176 (2020), p.~107685.

\bibitem{chen2012new}
{\sc H.~Chen and B.~Zeng}, {\em New transforms tightly bounded by {DCT} and
  {KLT}}, IEEE Signal Processing Letters, 19 (2012), pp.~344--347.

\bibitem{chen2019hardware}
{\sc J.~Chen, S.~Liu, G.~Deng, and S.~Rahardja}, {\em Hardware efficient
  integer discrete cosine transform for efficient image/video compression},
  IEEE Access, 7 (2019), pp.~152635--152645.

\bibitem{chen1977fast}
{\sc W.-H. Chen, C.~Smith, and S.~Fralick}, {\em A fast computational algorithm
  for the discrete cosine transform}, IEEE Transactions on Communications, 25
  (1977), p.~1004–1009.

\bibitem{cintra2011dct}
{\sc R.~J. Cintra and F.~M. Bayer}, {\em A {DCT} approximation for image
  compression}, IEEE Signal Processing Letters, 18 (2011), p.~579–582.

\bibitem{cintra2014low}
{\sc R.~J. Cintra, F.~M. Bayer, and C.~Tablada}, {\em Low-complexity 8-point
  {DCT} approximations based on integer functions}, Signal Processing, 99
  (2014), p.~201–214.

\bibitem{coelho2021scaling}
{\sc D.~F. Coelho, R.~J. Cintra, A.~Madanayake, and S.~M. Perera}, {\em
  Low-complexity scaling methods for \mbox{DCT-II} approximations}, IEEE
  Transactions on Signal Processing,  (2021), pp.~4557--4566.

\bibitem{da2017multiplierless}
{\sc T.~L. da~Silveira, R.~S. Oliveira, F.~M. Bayer, R.~J. Cintra, and
  A.~Madanayake}, {\em Multiplierless 16-point {DCT} approximation for
  low-complexity image and video coding}, Signal, Image and Video Processing,
  11 (2017), p.~227–233.

\bibitem{fan2019signal}
{\sc K.~Fan, R.~Wang, W.~Lin, L.-Y. Duan, and W.~Gao}, {\em Signal-independent
  separable {KLT} by offline training for video coding}, IEEE Access, 7 (2019),
  p.~33087–33093.

\bibitem{flach2012machine}
{\sc P.~Flach}, {\em Machine learning: the art and science of algorithms that
  make sense of data}, Cambridge University Press, 2012.

\bibitem{geetha2020hybrid}
{\sc V.~Geetha, V.~Anbumani, G.~Murugesan, and S.~Gomathi}, {\em Hybrid optimal
  algorithm-based {2D} discrete wavelet transform for image compression using
  fractional {KCA}}, Multimedia Systems, 26 (2020), pp.~687--702.

\bibitem{gonzalez2002digital}
{\sc R.~C. Gonzalez, R.~E. Woods, et~al.}, {\em Digital image processing},
  Prentice hall, Upper Saddle River, NJ, 2002.

\bibitem{hao2003reversible}
{\sc P.~Hao and Q.~Shi}, {\em Reversible integer {KLT} for
  progressive-to-lossless compression of multiple component images}, in
  Proceedings 2003 International Conference on Image Processing, vol.~1, IEEE,
  2003, pp.~I--633.

\bibitem{hartigan1979algorithm}
{\sc J.~A. Hartigan and M.~A. Wong}, {\em Algorithm {AS} 136: A $k$-means
  clustering algorithm}, Journal of the Royal Statistical Society. Series C
  (Applied Statistics), 28 (1979), p.~100–108.

\bibitem{harville1997trace}
{\sc D.~A. Harville}, {\em Trace of a (square) matrix}, in Matrix Algebra From
  a Statistician's Perspective, Springer, New York, 1997, p.~49–53.

\bibitem{haweel2001new}
{\sc T.~I. Haweel}, {\em A new square wave transform based on the {DCT}},
  Signal Processing, 81 (2001), p.~2309–2319.

\bibitem{huang2019deterministic}
{\sc J.~Huang, T.~N. Kumar, H.~A. Almurib, and F.~Lombardi}, {\em A
  deterministic low-complexity approximate (multiplier-less) technique for
  {DCT} computation}, IEEE Transactions on Circuits and Systems I: Regular
  Papers, 66 (2019), pp.~3001--3014.

\bibitem{Huynh-Thu2008Scope}
{\sc Q.~Huynh-Thu and M.~Ghanbari}, {\em Scope of validity of {PSNR} in
  image/video quality assessment}, Electronics Letters, 44 (2008),
  p.~800–801.

\bibitem{jain1976fast}
{\sc A.~K. Jain}, {\em A fast {K}arhunen-{L}oève transform for a class of
  random processes}, IEEE Transactions on Communications, 24 (1976),
  p.~1023–1029.

\bibitem{jayakumar2020karhunen}
{\sc R.~Jayakumar and S.~Dhandapani}, {\em Karhunen {L}o{\`e}ve transform with
  adaptive dictionary learning for coherent and random noise attenuation in
  seismic data}, S{\=a}dhan{\=a}, 45 (2020), pp.~1--13.

\bibitem{jridi2015generalized}
{\sc M.~Jridi, A.~Alfalou, and P.~K. Meher}, {\em A generalized algorithm and
  reconfigurable architecture for efficient and scalable orthogonal
  approximation of {DCT}}, IEEE Transactions on Circuits and Systems I: Regular
  Papers, 62 (2015), p.~449–457.

\bibitem{karhunen1947under}
{\sc K.~Karhunen}, {\em Under lineare methoden in der wahr
  scheinlichkeitsrechnung}, Annales Academiae Scientiarun Fennicae Series A1:
  Mathematia Physica, 47 (1947).

\bibitem{katto1992short}
{\sc J.~Katto, K.~Komatsu, and Y.~Yasuda}, {\em Short-tap and linear-phase {PR}
  filter banks for subband coding of images}, in Visual Communications and
  Image Processing'92, vol.~1818, International Society for Optics and
  Photonics, 1992, p.~735–747.

\bibitem{lan1993fast}
{\sc L.-S. Lan and I.~S. Reed}, {\em Fast approximate {K}arhunen-{L}o{\`e}ve
  transform with applications to digital image coding}, in Visual
  Communications and Image Processing'93, vol.~2094, International Society for
  Optics and Photonics, 1993, pp.~444--455.

\bibitem{lan1994improved}
\leavevmode\vrule height 2pt depth -1.6pt width 23pt, {\em An improved {JPEG}
  image coder using the adaptive fast approximate {K}arhunen-{L}o{\`e}ve
  transform ({AKLT})}, in Proceedings of ICSIPNN'94. International Conference
  on Speech, Image Processing and Neural Networks, IEEE, 1994, pp.~160--163.

\bibitem{loeffler1989practical}
{\sc C.~Loeffler, A.~Ligtenberg, and G.~S. Moschytz}, {\em Practical fast {1-D}
  {DCT} algorithms with 11 multiplications}, in Acoustics, Speech, and Signal
  Processing, 1989. ICASSP-89., 1989 International Conference on, IEEE, 1989,
  p.~988–991.

\bibitem{loeve1948functions}
{\sc M.~Loève}, {\em Fonctions aléatoires de second ordre}, Processus
  {S}tochastique et {M}ouvement {B}rownien,  (1948), p.~366–420.

\bibitem{mefoued2023improving}
{\sc A.~Mefoued, N.~Kouadria, S.~Harize, and N.~Doghmane}, {\em Improving image
  encoding quality with a low-complexity dct approximation using 14 additions},
  Journal of Real-Time Image Processing, 20 (2023), p.~58.

\bibitem{ochoa2019discrete}
{\sc H.~Ochoa-Dominguez and K.~R. Rao}, {\em Discrete Cosine Transform}, CRC
  Press, Boca Raton, 2019.

\bibitem{oliveira2015discrete}
{\sc P.~A. Oliveira, R.~J. Cintra, F.~M. Bayer, S.~Kulasekera, and
  A.~Madanayake}, {\em A discrete {T}chebichef transform approximation for
  image and video coding}, IEEE Signal Processing Letters, 22 (2015),
  p.~1137–1141.

\bibitem{oliveira2019low}
{\sc R.~S. Oliveira, R.~J. Cintra, F.~M. Bayer, T.~L. da~Silveira,
  A.~Madanayake, and A.~Leite}, {\em Low-complexity 8-point {DCT} approximation
  based on angle similarity for image and video coding}, Multidimensional
  Systems and Signal Processing, 30 (2019), p.~1363–1394.

\bibitem{pirooz1998new}
{\sc A.~Pirooz and I.~Reed}, {\em A new approximate {K}arhunen-{L}o{\`e}ve
  transform for data compression}, in Conference Record of Thirty-Second
  Asilomar Conference on Signals, Systems and Computers, vol.~2, IEEE, 1998,
  pp.~1471--1475.

\bibitem{plonka2004global}
{\sc G.~Plonka}, {\em A global method for invertible integer {DCT} and integer
  wavelet algorithms}, Applied and Computational Harmonic Analysis, 16 (2004),
  p.~90–110.

\bibitem{potluri2014improved}
{\sc U.~S. Potluri, A.~Madanayake, R.~J. Cintra, F.~M. Bayer, S.~Kulasekera,
  and A.~Edirisuriya}, {\em Improved 8-point approximate {DCT} for image and
  video compression requiring only 14 additions}, IEEE Transactions on Circuits
  and Systems I: Regular Papers, 61 (2014), p.~1727–1740.

\bibitem{pourazad2012hevc}
{\sc M.~T. Pourazad, C.~Doutre, M.~Azimi, and P.~Nasiopoulos}, {\em {HEVC}: The
  new gold standard for video compression: How does {HEVC} compare with
  \mbox{{H}.264}/{AVC}?}, IEEE Consumer Electronics Magazine, 1 (2012),
  p.~36–46.

\bibitem{puchala2021approximate}
{\sc D.~Puchala}, {\em Approximate calculation of $8$-point {DCT} for various
  scenarios of practical applications}, EURASIP Journal on Image and Video
  Processing, 2021 (2021), pp.~1--34.

\bibitem{Puri2004}
{\sc A.~Puri}, {\em Video coding using the {H.264/MPEG}-4 {AVC} compression
  standard}, Signal Processing: Image Communication, 19 (2004).

\bibitem{radunz2021data}
{\sc A.~Rad{\"u}nz, T.~da~Silveira, F.~Bayer, and R.~Cintra}, {\em
  Data-independent low-complexity {KLT} approximations for image and video
  coding}, Signal Processing: Image Communication,  (2021).

\bibitem{radunz2021low}
{\sc A.~P. Rad{\"u}nz, F.~M. Bayer, and R.~J. Cintra}, {\em Low-complexity
  rounded {KLT} approximation for image compression}, Journal of Real-Time
  Image Processing,  (2021), pp.~1--11.

\bibitem{radunz2023extensions}
{\sc A.~P. Rad{\"u}nz, L.~Portella, R.~Oliveira, F.~M. Bayer, and R.~J.
  Cintra}, {\em Extensions on low-complexity dct approximations for larger
  blocklengths based on minimal angle similarity}, Journal of Signal Processing
  Systems, 95 (2023), pp.~495--516.

\bibitem{rao1990yip}
{\sc K.~R. Rao and P.~C. Yip}, {\em Discrete cosine transform, algorithm,
  advantage and applications}, New York: Academic,  (1990).

\bibitem{rao2000transform}
\leavevmode\vrule height 2pt depth -1.6pt width 23pt, {\em The transform and
  data compression handbook}, vol.~1, CRC press, Boca Raton, FL, 2000.

\bibitem{ray1970further}
{\sc W.~Ray and R.~Driver}, {\em Further decomposition of the
  {K}arhunen-{L}o\`eve series representation of a stationary random process},
  IEEE Transactions on Information Theory, 16 (1970), p.~663–668.

\bibitem{reed1994fast}
{\sc I.~S. Reed and L.-S. Lan}, {\em A fast approximate {K}arhunen-{L}oève
  transform {(AKLT)} for data compression}, Journal of Visual Communication and
  Image Representation, 5 (1994), p.~304–316.

\bibitem{Safiri1996}
{\sc H.~Safiri, M.~Ahmadi, G.~A. Jullien, and V.~S. Dimitrov}, {\em Design and
  {FPGA} implementation of systolic {FIR} filters using the fermat number
  {ALU}}, in Asilomar Conference on Signals, Systems and Computers, vol.~2,
  1996, pp.~1052--1056.

\bibitem{salomon2004data}
{\sc D.~Salomon}, {\em Data compression: the complete reference}, Springer
  Science {\&} Business Media, New York, 2004.

\bibitem{seber2008matrix}
{\sc G.~A. Seber}, {\em A matrix handbook for statisticians}, vol.~15, John
  Wiley {\&} Sons, New Jersey, 2008.

\bibitem{singhadia2019novel}
{\sc A.~Singhadia, P.~Bante, and I.~Chakrabarti}, {\em A novel algorithmic
  approach for efficient realization of {2-D}-{DCT} architecture for {HEVC}},
  IEEE Transactions on Consumer Electronics, 65 (2019), pp.~264--273.

\bibitem{sipi2005usc}
{\sc U.~SIPI}, {\em The {USC-SIPI} image database}, 1977.

\bibitem{sole2009joint}
{\sc J.~Sole, P.~Yin, Y.~Zheng, and C.~Gomila}, {\em Joint sparsity-based
  optimization of a set of orthonormal {2-D} separable block transforms}, in
  2009 16th IEEE International Conference on Image Processing (ICIP), IEEE,
  2009, pp.~9--12.

\bibitem{suzuki2010integer}
{\sc T.~Suzuki and M.~Ikehara}, {\em Integer {DCT} based on direct-lifting of
  {DCT-IDCT} for lossless-to-lossy image coding}, IEEE Transactions on Image
  Processing, 19 (2010), p.~2958–2965.

\bibitem{tablada2017dct}
{\sc C.~Tablada, T.~L.~T. da~Silveira, R.~J. Cintra, and F.~M. Bayer}, {\em
  {DCT} approximations based on {C}hen’s factorization}, Signal Processing:
  Image Communication, 58 (2017), p.~14–23.

\bibitem{nikara2001unified}
{\sc J.~Takala and J.~Nikara}, {\em Unified pipeline architecture for discrete
  sine and cosine transforms of type {IV}}, in Proceedings of the 3rd
  Internacional Conference on Information Communication and Signal Processing,
  2001.

\bibitem{thomakos2016smoothing}
{\sc D.~Thomakos}, {\em Smoothing non-stationary time series using the discrete
  cosine transform}, Journal of Systems Science and Complexity, 29 (2016),
  pp.~382--404.

\bibitem{wallace1992jpeg}
{\sc G.~K. Wallace}, {\em The {JPEG} still picture compression standard}, IEEE
  Transactions on Consumer Electronics, 38 (1992), p.~xviii–xxxiv.

\bibitem{wang2009mean}
{\sc Z.~Wang and A.~C. Bovik}, {\em Mean squared error: Love it or leave it? a
  new look at signal fidelity measures}, IEEE Signal Processing Magazine, 26
  (2009), pp.~98--117.

\bibitem{wang2004image}
{\sc Z.~Wang, A.~C. Bovik, H.~R. Sheikh, and E.~P. Simoncelli}, {\em Image
  quality assessment: from error visibility to structural similarity}, IEEE
  Transactions on Image Processing, 13 (2004), p.~600–612.

\bibitem{wongsawat2006integer}
{\sc Y.~Wongsawat, S.~Oraintara, and K.~R. Rao}, {\em Integer sub-optimal
  {K}arhunen-{L}o{\`e}ve transform for multi-channel lossless {EEG}
  compression}, in 2006 14th European Signal Processing Conference, IEEE, 2006,
  pp.~1--5.

\bibitem{yang2020blind}
{\sc C.~Yang, X.~Zhang, P.~An, L.~Shen, and C.-C.~J. Kuo}, {\em Blind image
  quality assessment based on multi-scale {KLT}}, IEEE Transactions on
  Multimedia,  (2020).

\bibitem{yanyun2004updating}
{\sc Q.~Yanyun, Z.~Nanning, L.~Cuihua, and Y.~Zejian}, {\em Updating algorithm
  for extracting the basis of {K}arhunen-{L}o{\`e}ve transform in nonzero mean
  data}, in Proceedings 7th International Conference on Signal Processing,
  2004. Proceedings. ICSP'04. 2004., vol.~2, IEEE, 2004, pp.~1403--1406.

\bibitem{zhang2020data}
{\sc X.~Zhang, S.~Kwong, and C.-C.~J. Kuo}, {\em Data-driven transform based
  compressed image quality assessment}, IEEE Transactions on Circuits and
  Systems for Video Technology,  (2020).

\bibitem{zidani2019low}
{\sc N.~Zidani, N.~Kouadria, N.~Doghmane, and S.~Harize}, {\em Low complexity
  pruned {DCT} approximation for image compression in wireless multimedia
  sensor networks}, in 2019 5th International Conference on Frontiers of Signal
  Processing ({ICFSP}), IEEE, 2019, pp.~26--30.

\end{thebibliography}
}

\end{document}